\PassOptionsToPackage{unicode}{hyperref}
\PassOptionsToPackage{hyphens}{url}
\documentclass[12pt]{article}
\usepackage{xcolor}
\usepackage[margin=1in]{geometry}
\usepackage{amsmath,amssymb}
\usepackage[T1]{fontenc}
\usepackage[utf8]{inputenc}
\usepackage{textcomp}
\usepackage{lmodern}
\usepackage{CJKutf8}
\IfFileExists{upquote.sty}{\usepackage{upquote}}{}
\IfFileExists{microtype.sty}{%
  \usepackage[]{microtype}
  \UseMicrotypeSet[protrusion]{basicmath}
}{}
\makeatletter
\@ifundefined{KOMAClassName}{%
  \IfFileExists{parskip.sty}{\usepackage{parskip}}{%
    \setlength{\parindent}{0pt}
    \setlength{\parskip}{6pt plus 2pt minus 1pt}}
}{\KOMAoptions{parskip=half}}
\makeatother
\usepackage{longtable,booktabs,array}
\usepackage{calc}
\usepackage{etoolbox}
\makeatletter
\patchcmd\longtable{\par}{\if@noskipsec\mbox{}\fi\par}{}{}
\makeatother
\IfFileExists{footnotehyper.sty}{\usepackage{footnotehyper}}{\usepackage{footnote}}
\makesavenoteenv{longtable}
\providecommand{\tightlist}{\setlength{\itemsep}{0pt}\setlength{\parskip}{0pt}}
\usepackage[]{natbib}
\usepackage{graphicx}
\usepackage{bookmark}
\IfFileExists{xurl.sty}{\usepackage{xurl}}{}
\hypersetup{hidelinks, pdfcreator={LaTeX via pandoc}}

\title{Agentic Social Affordance Framework (ASAF): Agent Identity Design as a Collaboration Interface in Multi-Agent Systems}

\author{
  Meng-Han Lee \\
  \textit{Independent Researcher, Taipei, Taiwan} \\
  \href{https://orcid.org/0009-0007-1685-0877}{ORCID: 0009-0007-1685-0877} \\
  \texttt{zaious.design@gmail.com}
}

\date{June 29, 2026 \quad|\quad Preprint \quad|\quad License: CC BY 4.0}

\begin{document}
\begin{CJK}{UTF8}{bsmi}

\maketitle

\begin{center}
\textit{Under review at Frontiers in Computer Science (Human-Media Interaction). This version has been revised in response to reviewer feedback.}
\end{center}
\medskip

\section{Abstract}\label{abstract}

As AI systems evolve from single conversational agents to complex
multi-agent architectures, a critical design dimension has been
overlooked: how the social identity of individual agents shapes human
behavior within the collaboration. This paper introduces the Agentic
Social Affordance Framework (ASAF), a theoretical framework that extends
Social Affordance theory into the context of multi-agent AI systems. We
propose that agent identity design functions not merely as a user
interface convention, but as a collaboration interface---structuring how
users perceive, approach, and engage with each agent, and thereby
influencing the quality of Human-Agent collaboration outcomes.
Specifically, ASAF adopts the analytical separability of the social
affordance layer and the engineering orchestration layer as a
\textbf{framing assumption}---an organizing distinction that structures
design analysis---rather than as a testable claim about
effect-independence. The layers' downstream effects can and do interact;
the framing assumption is useful because it identifies a class of design
considerations (cognitive posture, role-schema activation, oversight
calibration) that is irreducible to engineering choices alone. ASAF
comprises three mechanisms: Identity Signaling, Behavioral Priming, and
Collaborative Governance, and specifies their boundary conditions
through a four-tier Identity Signal Fidelity Spectrum and an
individual-difference moderating variable (anthropomorphizing
vs.~instrumentalizing cognitive style). We situate ASAF in relation to
existing affordance theory (Hutchby, 2001), the CASA paradigm (Gambino
et al., 2020), and classical multi-agent systems research (Wooldridge \&
Jennings, 1995), identifying a directional reversal: where classical MAS
used roles, norms, and coordination to constrain autonomous agents, ASAF
applies the same organizational vocabulary to structure the cognition
and oversight of human operators who remain in the loop. ASAF positions
social affordance design as a first-class design responsibility that
engineering orchestration cannot subsume, with concrete implications for
multi-agent system design. We outline directions for future empirical
validation, including a factorial design for characterizing the
empirical interaction surface between the social affordance and
engineering orchestration layers.

\textbf{Keywords:} Social Affordance, Multi-Agent Systems, Human-Agent
Interaction, Agent Identity Design, Identity Signal, LLM Agent,
Human-in-the-Loop, Cognitive Scaffolding

\begin{center}\rule{0.5\linewidth}{0.5pt}\end{center}

\section{1. Introduction}\label{introduction}

Recent advancements in multi-agent artificial intelligence have driven
increased research on orchestrational efficiency, token optimization,
and inter-agent communication protocols (e.g., Hong et al., 2023). While
this engineering-centric discourse has successfully established the
technical scaffolding for collaborative AI, it inherently marginalizes
the human-computer interaction (HCI) dimension. Modern multi-agent
systems are rarely closed causal loops; they exist primarily in
Human-in-the-Loop (HitL) configurations where human operators must
direct, parse, and trust the outputs of highly specialized, distinct
\emph{node agents}---individual agents occupying defined role-bearing
positions within the orchestration topology, each carrying a declared
identity and a designated channel of inter-agent communication.

The impetus for this conceptual framework emerged directly from the
author's industrial deployment challenges rather than isolated
laboratory experiments. During the rapid proliferation of orchestration
engines (e.g., LangGraph, AutoGen), the developer community
overwhelmingly anthropomorphized agent workflows into corporate team
structures (detailed in Section 2.3), focusing intensely on which
execution engine to adopt. However, when scaling a custom collaborative
system from a single conversational assistant to a scalable multi-agent
personal productivity architecture comprising dozens of specialized
nodes (the \emph{ChronicleCore} system detailed in Section 4), a
critical reality became apparent: the true bottleneck is rarely the
underlying orchestration framework. Instead, the fundamental friction
point is \emph{Context Governance}: the human operator's capacity to
navigate and interact with disparate computational nodes without
inducing cognitive collapse or systemic persona drift.

This realization bridges the gap to established literature. Historical
meta-analyses in human-robot interaction suggest that human trust and
collaborative effectiveness are significantly influenced by an agent's
attribute-based design and perceived role (Hancock et al., 2011). In the
context of large language models, this dynamic manifests through
``role-play,'' where system prompts cast the model into specific
functional and social identities, establishing boundaries for its
computational behavior (Shanahan et al., 2023). However, the
implications of these role-playing dynamics observed within a cohesive
multi-agent topology remain critically under-theorized.

This paper argues that the social identity of individual agents within a
network functions not merely as a cosmetic persona, but as a structural
collaboration interface. Anthropomorphic architecture in multi-agent
systems is neither an emotional user-experience feature nor an arbitrary
engineering choice; rather, it is an inevitable design response to human
cognitive architecture confronting systems whose scale exceeds
unassisted working memory. We adopt working memory capacity estimates
(Miller, 1956; Cowan, 2001) as an approximate problem-scale heuristic.
While Cowan's measure concerns simultaneous item retention in short-term
memory rather than active cross-session navigation, it establishes the
cognitive order of magnitude at which unstructured information
management becomes untenable; the cognitive offloading literature (Risko
\& Gilbert, 2016) demonstrates that when task demands approach such
capacity limits, users spontaneously seek external scaffolding
structures. ASAF posits that at this approximate scale (approximately
4--7 unstructured functional profiles, depending on chunking strategy
and individual differences), the application of social schemas
transforms from a superficial preference into a structural necessity
(i.e., an indispensable design consideration, not a guaranteed effect
for all users; effect magnitude remains moderated by individual
cognitive style, as specified in H1).

ASAF distinguishes itself from existing theories (such as CASA,
classical role theory, and distributed cognition) through an
\textbf{analytical framing assumption}: the social affordance layer is
treated as a class of design considerations that is \emph{not reducible}
to engineering orchestration choices. We adopt this framing
assumption---rather than claim it as a testable thesis about
effect-independence---because the considerations a designer addresses
when configuring agent identity (what cognitive posture to signal, what
role schemata to activate, what oversight calibration to invite) form a
coherent design vocabulary that engineering decisions about
orchestration protocols, tool exposure, or inter-agent message passing
do not exhaust. The framing assumption is useful because it makes
visible a design surface that engineering-centric accounts otherwise
leave implicit. The substantive characterization of this design surface
is developed below; its empirical consequences---how identity design
effects manifest, interact with engineering effects, and are moderated
by user cognitive style---are addressed in H1--H3b (Section 5).

To clarify the boundary with CASA (Reeves \& Nass, 1996): CASA's core
prediction---that humans automatically apply social norms to
computers---can account for the basic Identity Signaling effect at the
dyadic level (i.e., a single user responding socially to a single
personified agent). However, CASA is structurally a dyadic theory; it
does not generate predictions about (a) the scaling threshold at which
social identity shifts from optional to structurally important
(approximately 4--7 agents; H1), (b) the differential oversight
calibration across agents occupying distinct topological roles within a
system---specifically, that the same agent identity elicits different
operator behavior depending on the surrounding agents' role
configuration (H3b, Section 5), or (c) the analytical framing of the
social design space as a class of considerations not reducible to
engineering orchestration choices. CASA-plus-role-schema accounts
(Gambino et al., 2020; Biddle, 1986) can generate the basic role-level
oversight differential (H3a), but no extension of CASA---including
aggregated or conformity-based variants (cf.~Song et al.,
2025)---generates the topology-level prediction in H3b, because CASA
lacks a structural vocabulary for agent-topology-level oversight
calibration. ASAF's independent contribution lies precisely in these
multi-agent, topology-level, and design-theoretic dimensions that fall
outside CASA's explanatory scope.

Under this framing assumption, the social identity design and
engineering orchestration design address distinct classes of
considerations: configuring agent identity (cognitive posture,
interaction norms, role-schema activation) draws on a different design
vocabulary than configuring capability (tool exposure, orchestration
protocol, message-passing scheme). Engineering restructuring does not
entail any particular social identity configuration, and redesigning
agent personas does not alter the engineering topology---a separability
that is observable at the level of design decisions even when the
layers' downstream effects interact. H1 (Section 5) tests one
consequence of this framing: whether social identity design carries
independent explanatory power beyond functional labeling. The broader
interaction surface between the two layers is addressed by the factorial
design discussed in Section 7. Consequently, treating multi-agent design
purely as a software engineering decomposition problem risks creating
structurally coherent but interactionally unreadable systems.

This paper makes the following contributions:

\begin{enumerate}
\def\labelenumi{\arabic{enumi}.}
\tightlist
\item
  We introduce the Agentic Social Affordance Framework (ASAF),
  distinguishing its predictive power by positioning anthropomorphic
  identity not as an empathetic illusion, but as a structurally
  important cognitive load-reduction interface for human operators
  managing systems that exceed working memory capacity (approximately
  4--7 agents, depending on chunking strategy and individual
  differences). Its effect magnitude is moderated by individual
  cognitive style (anthropomorphizing vs.~instrumentalizing tendency),
  as specified in H1.
\item
  We identify and define three core mechanisms through which ASAF
  operates: Identity Signaling, Behavioral Priming, and Collaborative
  Governance.
\item
  We situate ASAF in relation to existing affordance theory (Gibson,
  1979; Norman, 1988) and automation reliance literature (Parasuraman \&
  Riley, 1997; Lee \& See, 2004).
\item
  We provide operational design boundaries, formally defining how
  individual differences in cognitive style (anthropomorphizing
  vs.~instrumentalizing) moderate the framework's predictive validity.
\item
  We identify an ontological displacement and directional reversal from
  classical multi-agent systems (Wooldridge \& Jennings, 1995): where
  classical MAS treated human operators---when considered at all---as
  just another type of agent assimilated into a framework presupposing
  autonomy, rationality, and stable utility, ASAF applies the same
  organizational vocabulary to structure the cognition of human
  operators---theorized as structurally distinct from the agents they
  oversee---in systems where agents lack the endogenous autonomy that
  the classical formalism required.
\end{enumerate}

\begin{center}\rule{0.5\linewidth}{0.5pt}\end{center}

\section{2. Background and Related
Work}\label{background-and-related-work}

\subsection{2.1 Affordance Theory: From Gibson to Social
Affordance}\label{affordance-theory-from-gibson-to-social-affordance}

The concept of affordance was introduced by Gibson (1979) as a
relational property between an actor and their environment: the action
possibilities that an environment offers to a particular actor.
Affordances are neither purely objective properties of objects nor
purely subjective constructions of the mind; they emerge from the
relationship between the two.

Norman (1988; 1999) operationalized affordance for design contexts,
distinguishing between real affordances (what an object actually allows)
and perceived affordances (what users believe it allows). Norman's
contribution established affordance as a central concept in
human-computer interaction, directing attention to how design choices
shape user perception and behavior.

Social Affordance extends this framework to the social dimensions of
technology. Where Gibson's affordances describe physical action
possibilities and Norman's describe functional ones, Social Affordances
describe the social interaction possibilities that a system's design
signals to its users (Hutchby, 2001). During the Web 2.0 era, this was
extensively applied to analyze how the spatial architecture of social
network sites shaped ``networked publics'' and structured human
connection (Boyd, 2010). Nagy and Neff (2015) advanced this domain by
introducing the concept of \emph{imagined affordances}, arguing that a
user's \emph{belief} about what a technology or algorithmic system can
do is just as determinant of their behavior as the technology's
hard-coded capabilities. This transitioned affordance theory from
analyzing static interfaces to analyzing socially constructed
expectations.

Recent sociological work has begun bridging this framework into
artificial intelligence. A critical conceptual precursor is Sundar's
(2020) framework for Human-AI Interaction, which builds on his earlier
Theory of Interactive Media Effects (TIME) and examines the tension
between human and machine agency. Sundar identifies the ``machine
heuristic''---where users apply mental shortcuts to evaluate AI outputs
based on the mere presence of a machine interface. While Sundar's
machine heuristic operates as a binary trigger (machine vs.~human),
Identity Signaling in multi-agent systems operates as a multi-valued,
role-differentiated signal: users do not merely respond to ``a machine''
but calibrate their behavior to \emph{which kind of social role} the
machine presents. This shift from binary agency detection to structured
role-based calibration is precisely the granularity gap that ASAF
addresses. Empirical studies on ChatGPT usage in specific cultural
contexts illustrate how perceived affordances in LLMs are shaped by user
interpretation practices rather than fixed machine properties alone
(Haqqu et al., 2025). Anthropomorphic design cues further trigger
parasocial relationship dynamics, modifying user behavior without
changing the underlying model parameters (Reeves \& Nass, 1996; Epley et
al., 2007). Together, these findings indicate that affordance theory
must evolve past simple ``machine heuristics'' toward interpreting the
explicit role schemata embedded in system design.

\subsection{2.2 The Chatbot Era: Single-Agent Social
Affordance}\label{the-chatbot-era-single-agent-social-affordance}

In the Chatbot era, Social Affordance operated through a single, often
ambiguous social identity. ChatGPT, Claude, and similar systems present
a unified interface that affords many things---assistance, conversation,
analysis, creativity---but commits to none. This ambiguity is a design
choice that maximizes perceived utility but minimizes Social Affordance
specificity. Users engage with these systems through general-purpose
interaction patterns, rarely adjusting their behavior based on a
perceived social role of the system.

The limitation is structural: with a single agent identity, Social
Affordance can only operate at the level of the whole system. There is
no mechanism for users to adjust their interaction style based on the
specific capability or social role of a particular component.

\subsection{2.3 The Evolution of Agentic Roles: The ``Virtual Team''
Wave and Its
Backlash}\label{the-evolution-of-agentic-roles-the-virtual-team-wave-and-its-backlash}

Multi-agent AI systems introduce a structurally distinct architecture,
replacing single conversational models with networks of specialized
agents. Beginning in late 2025 and accelerating through the first
quarter of 2026, a rapid resurgence of anthropomorphic multi-agent
architectures emerged, what we term here the \textbf{Virtual Team Wave}.
Developers returned to structuring agents as corporate role hierarchies,
and in some cases, historical administrative metaphors. Three
representative open-source implementations collectively accumulated
\textbf{169,732 GitHub stars} within this window, surfacing an apparent
industry-wide appetite for persona-structured agent fleets.\footnote{Repository
  metrics retrieved via the GitHub REST API (GET
  /repos/\{owner\}/\{repo\}) on 2026-04-16. Aggregate at retrieval:
  169,732 stars, 24,954 forks across the three repositories. These
  figures aggregate three distinct repositories; actual unique observer
  count is lower due to overlap among early-adopter communities who star
  multiple related projects---a sampling bias inherent to GitHub stars
  as a social signal inflated by high-profile distribution events (e.g.,
  curator endorsements). The metric is cited as a directional indicator
  of industry interest, not as a precise adoption measure. Independent
  of open-source enthusiasm, enterprise deployments in Q1--Q2 2026 have
  begun embedding agent identity into infrastructure layers (e.g.,
  Microsoft Entra Agent ID, Google Agent Registry, AWS Bedrock
  AgentCore), suggesting that role-structured agent architectures are
  transitioning from experimental practice to production infrastructure.}

These implementations illustrate both the scale of the trend and the
cross-cultural nature of the underlying design impulse. \emph{The
Agency} (Sitarzewski, 2025), which seeded the wave in October 2025,
organizes 147 agents into corporate-style divisions (engineering,
design, marketing, product), each agent written as ``a specialized
expert with personality, processes, and proven deliverables.''
\emph{gstack} (Tan, 2026), authored by the President of Y Combinator and
released in March 2026, operationalizes a single builder's team as ``23
opinionated tools that serve as CEO, Designer, Eng Manager, Release
Manager, Doc Engineer, and QA,'' accumulating over 73,000 stars within
its first five weeks. \emph{Edict} (cft0808, 2026; the maintainer
publishes under this GitHub handle) departs from the Silicon Valley
corporate metaphor entirely, adopting the \textbf{Three Departments and
Six Ministries} (三省六部) system of Tang Dynasty governance (ca. 618
CE). It distributes agent roles across the Secretariat, the Chancellery,
and the Department of State Affairs, operating under a strict
inter-agent permission matrix. That the same underlying design impulse
surfaces in both Silicon Valley startup vernacular and 1,400-year-old
administrative structure suggests that this anthropomorphization is not
a local stylistic preference but a recurring cognitive response to the
orchestration demands of multi-agent systems. At the commercial
frontier, xAI's Grok 4.20 (released February 2026) deploys four named
specialist agents as its default inference architecture (xAI,
2026).\footnote{xAI's Grok 4.20 (released February 17, 2026) markets
  ``four expert-mode AI agents'' as a core feature of its paid SuperGrok
  tier (grok.com). The individual agent names---including Benjamin
  (mathematical reasoning) and Lucas (contrarian analysis)---are
  observable within the paid product interface (SuperGrok subscription
  required).} This structural shift towards roleplaying clusters is not
an isolated design anomaly but an emergent standard practice.
Qualitative studies of early adopters confirm that practitioners
organically conceptualize and interact with complex LLM pipelines as
``teams'' of specialized, task-based collaborators rather than mere
software threads (Naik et al., 2025).

However, while Naik et al.~(2025) establish \emph{that} early adopters
default to this team-oriented mental model, the underlying mechanism
shaping interaction remains untheorized. Empirical work concurrently
demonstrates that groups of AI agents can function as cohesive social
groups, exerting normative social pressure on users through mechanisms
analogous to human group dynamics (Song et al., 2025).

Recent empirical work has demonstrated that specific agent properties
(e.g., group presence (Song et al., 2025), verbalized uncertainty (Xu et
al., 2025), and confidence calibration (Li et al., 2024)) exert
measurable influence on human collaboration behavior. However, these
findings have yet to be integrated under a unifying design-level
framework. ASAF offers such an integration by treating each of these
properties as instances of Social Affordance signaling.

\subsection{2.3.1 Situating ASAF Against Classical Multi-Agent Systems
and
CASA}\label{situating-asaf-against-classical-multi-agent-systems-and-casa}

The organizational vocabulary that ASAF deploys---roles, norms,
coordination, governance---is not new; it descends directly from
classical multi-agent systems (MAS) research (Wooldridge \& Jennings,
1995; Smith, 1980) and the normative MAS tradition (Dignum, 2004). In
classical MAS, agents were assumed to possess endogenous
autonomy---entities capable of independently forming intentions,
commitments, and action plans (Rao \& Georgeff, 1995). Roles, norms,
coordination protocols, and filtering mechanisms were therefore designed
to constrain these autonomous agents into coherent collective behavior.

Modern LLM-based agents invert this assumption. They exhibit strong
generative capability while their coherence as persistent role-occupying
agents is externally supplied through orchestration, persistent memory,
and capability boundaries. This externalization is not merely a function
of current model limitations---it is also a deliberate architectural
choice for reasons of governability, auditability, and predictability,
and remains a viable design pattern independent of underlying model
capability. ASAF's predictive scope is bounded to architectures where
this externalization is the operative condition (Tiers 1-3 of the
Fidelity Spectrum, Section 2.3; deployment tier reflects operational
requirements rather than technological maturity). When classical MAS
considered human operators at all, they were assimilated as one type of
agent within the same formal framework---endowed with the same autonomy
assumptions, communication protocols, and rationality requirements as
artificial agents. The emergence of Human-Agent Interaction as a
distinct research tradition---rather than subsuming human interaction
within MAS's agent formalism---itself evidences this assimilation gap;
qualitative studies of early adopters further confirm that practitioners
organically conceptualize LLM pipelines as teams of specialized
collaborators rather than mere software threads (Naik et al., 2025), a
conceptualization that classical MAS's formalism was not designed to
accommodate.

Emerging critiques of LLM-based MAS have begun to identify this
ontological displacement from empirical, information-theoretic, and
cognitive grounds. ASAF does not rehearse these critiques. Rather, it
identifies their shared consequence at the cognitive layer: when the
agent ontology presupposed by classical MAS fails to transfer, the human
operator---systematically excluded from classical MAS theory or
assimilated as just another agent---becomes the system's actual
coordination bottleneck. ASAF therefore retains the MAS organizational
vocabulary---roles, norms, coordination, governance---but inverts its
directional target. This reversal---from agent-facing constraint to
human-facing cognitive interface---is what distinguishes ASAF from
classical MAS re-description.\footnote{La Malfa et al.~(2025), including
  Wooldridge, diagnose a similar ontological mismatch but prescribe
  closer alignment with classical MAS principles---a prescription ASAF
  does not share, as it treats the ontological gap not as a deviation to
  be corrected but as a structural feature of LLM-based agents that
  demands a new theoretical framework rather than stricter adherence to
  the old one.}

ASAF must also be distinguished from the Computers Are Social Actors
(CASA) paradigm (Reeves \& Nass, 1996) and its multi-source extensions
(Gambino, Fox, \& Ratan, 2020). CASA and its extensions treat social
responses to machines as automatic and mindless---a cognitive default
that occurs regardless of design intent. Gambino et al.~(2020) extend
this to media agents, arguing that users develop specific scripts for
human-media agent interaction that are, like human-human scripts,
applied mindlessly. ASAF operates in a condition that CASA did not
anticipate: LLM-based agents whose conversational performance is
sufficiently capable that role-based interaction is not a cognitive bias
to be explained away, but a functionally productive design strategy to
be deliberately engineered. In this condition, the question shifts from
``why do users treat machines as social actors?'' (CASA) to ``how should
designers structure agent identity to optimize human-agent
collaboration?'' (ASAF). Furthermore, ASAF's predictions operate at the
system topology level---the same agent identity elicits different
operator behavior depending on the topological role configuration of
surrounding agents (H3b, Section 5)---a structural property that dyadic
frameworks, including multi-source CASA extensions, do not generate.

ASAF must also be distinguished from the human-agent teaming tradition
(mixed-initiative interaction, coactive design, and adjustable
autonomy). These theories address how a human and one or more agents
partition initiative and dynamically adjust autonomy during task
execution; their basic unit is the human--agent relationship, whether
one agent or several under a single operator's supervision. ASAF
addresses a structurally different object: how agent identity design
structures the operator's cognitive load and differential oversight
across a multi-agent fleet. Crucially, the teaming unit---however many
agents it scales to---does not represent how the agents' identities are
configured relative to one another, because its formalism models the
human--agent link rather than the inter-agent identity topology. The two
frameworks are therefore not mutually exclusive; they describe distinct
layers of human--agent interaction (autonomy partition within
human--agent links vs.~identity-driven oversight across the agent
topology) and may co-apply within the same system. The empirical
signature isolating ASAF's layer is H3b (Section 5): the same agent
identity elicits different operator oversight as a function of the
surrounding agents' identity configuration---a topology-level prediction
that frameworks built on the human--agent link, including the teaming
tradition, do not structurally generate.

However, within the same window a fierce \textbf{Anti-Roleplay Backlash}
emerged in parallel. The developer community identified a fatal
architectural flaw in the ``virtual team'' model: \emph{information
decay during transmission}. Unlike humans who compensate for
communication gaps with shared cultural context, LLMs passing
synthesized documents down a rigid assembly line suffer from continuous
context loss and ``reasoning drift.'' Artificially imposing strict role
boundaries restricts the model's inherent cross-domain capabilities,
creating false walls precisely where valuable edge-case reasoning tends
to occur.

As a result, engineering consensus has swung violently away from
persona-driven pipelines. Leading infrastructure providers now
explicitly warn against fragmented agent architectures (e.g., Cognition
AI's critique of multi-agent fragility; Yan, 2025), prioritizing pure
deterministic paradigms instead. Frontier labs such as Anthropic
(2025a), for instance, have publicly advocated ``Context Engineering''
and the use of explicit shared state files (e.g., persistent
\texttt{progress.txt} or Git-based logs) over sequential roleplay. In
this current paradigm, agent identity is widely dismissed as a
dangerous, anthropomorphic illusion.

ASAF's mechanisms presuppose that agent identity signals remain
perceivable to the human operator across interactions. To contextualize
ASAF's scope against engineering critiques of context decay (e.g., Yan,
2025), we structure agent identity persistence along a four-tier design
spectrum:

\begin{enumerate}
\def\labelenumi{\arabic{enumi}.}
\tightlist
\item
  \textbf{Pure Prompt Injection}: Personas are defined per-session with
  no persistence, rendering them highly susceptible to reasoning drift.
\item
  \textbf{Prompt with Persistent Memory}: Personas span sessions via
  memory access, but lack structured cross-agent auditing.
\item
  \textbf{Structured Identity Enforcement}: Agents possess persistent
  structural modules, hard-gated orchestration layers, and cross-agent
  validation mechanisms (e.g., the ChronicleCore architecture detailed
  in Section 4).
\item
  \textbf{Model-Driven Routing}: Specialist routing occurs natively at
  inference-time and cannot be bypassed by the user (e.g., Grok 4.20).
\end{enumerate}

\begin{figure}[!htbp]
\centering
\includegraphics[width=\textwidth]{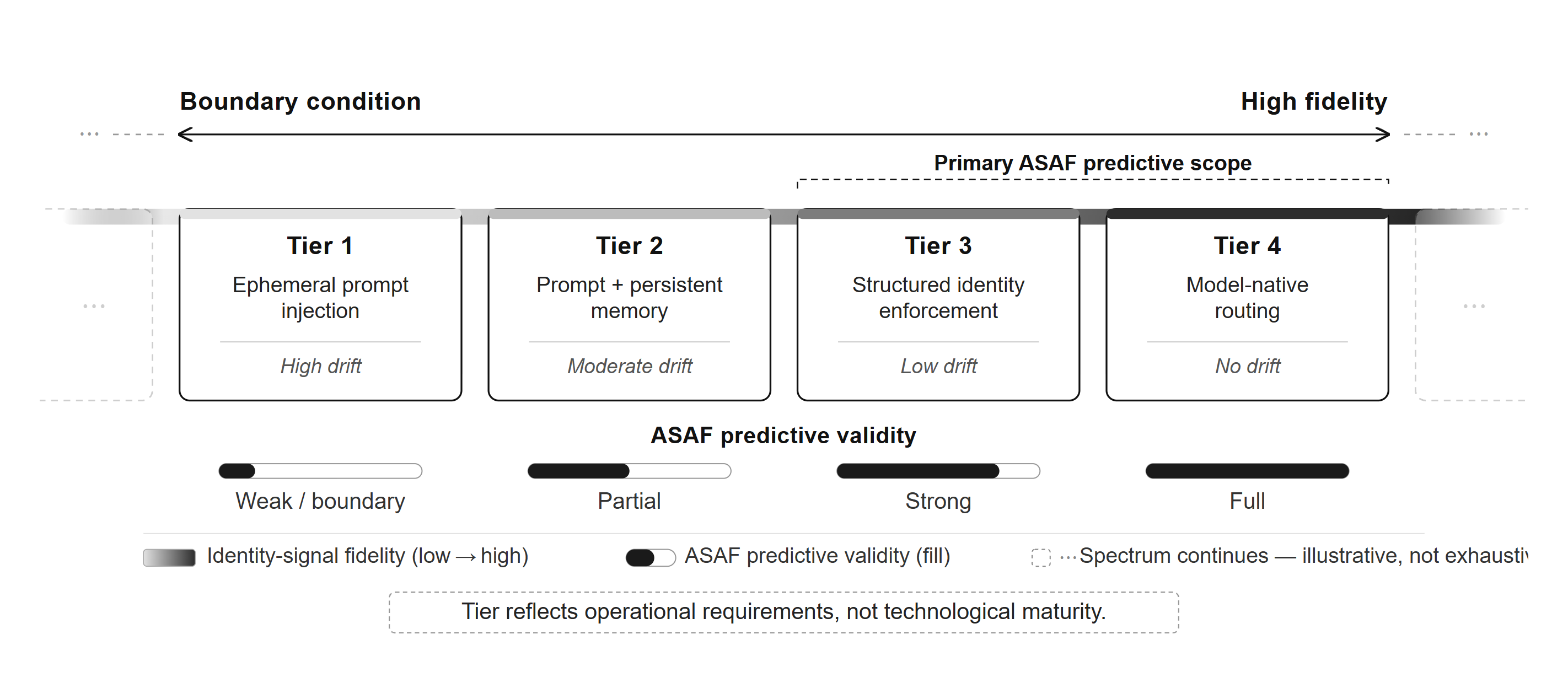}
\caption{The Identity Signal Fidelity Spectrum. ASAF's predictive validity scales with the structural enforcement of agent identity signals. Tier 1 (ephemeral prompt injection) represents a boundary condition where identity drift degrades framework predictions proportionally. Tiers 3--4, where identity signals are structurally enforced or model-native, constitute the primary scope of ASAF's strongest empirical predictions. The four tiers are illustrative rather than exhaustive---sampled points along a continuous spectrum (indicated by the fading band and open ends), not a closed enumeration. The spectrum is not a developmental trajectory---deployment tier is determined by operational requirements, not technological maturity.}
\label{fig:asaf-figure1}
\end{figure}

ASAF scopes its strongest predictions to systems operating at Tier 3 and
4, where Social Affordances are structurally enforced. The framework's
predictions apply partially to Tier 2, while Tier 1 serves as a boundary
condition where framework effectiveness degrades proportionally with
identity signal fidelity. Yan's (2025) critique accurately describes the
structural fragility inherent to Tier 1 and 2 deployments. However, his
critique targets the engineering layer (context fragility), while ASAF
operates on the cognitive layer (how human operators navigate deployed
architecture). These are analytically separable concerns: configuring
engineering robustness and configuring agent identity draw on distinct
design vocabularies, even when their downstream effects interact
(formalized as the analytical framing assumption in Section 3.1). The
appropriate tier for any deployment is determined by operational
requirements and risk tolerance, not by technological epoch. ASAF's
predictive validity scales with identity signal fidelity across this
spectrum---its mechanisms remain conceptually applicable at any tier,
with strongest empirical predictions at Tier 3-4 where structural
enforcement removes identity drift as a confound.

\subsection{2.4 The Cognitive Interface
Imperative}\label{the-cognitive-interface-imperative}

Building upon early theories of ``cognitive technologies'' that
reorganize human mental functioning (Pea, 1985; Salomon, 1993), and on
the growing literature on cognitive offloading in everyday tool use
(Risko \& Gilbert, 2016), we conceptually define a well-designed
multi-agent fleet as an external cognitive scaffold. A system whose
structure mirrors how the human architect's brain naturally decomposes
problems externalizes operations that would otherwise exhaust working
memory. When each agent carries a legible social identity, this
scaffolding effect becomes structured and predictable rather than
accidental (Amershi et al., 2019). The potential for deliberate
Human-Agent cognitive synergy of this kind remains largely absent from
current orchestration literature.

Reducing multi-agent architectures to purely functional software
pipelines ignores the true operational bottleneck of Human-in-the-Loop
systems. The ultimate hardware ceiling of any highly scaled agentic
system is not GPU capacity or API rate limits; it is the \emph{human
operator's own cognitive architecture}.

Foundational HCI research is unambiguous here: effective system design
must align with human mental models, not just maximize output accuracy
(Bansal et al., 2019). In the context of LLMs, Shanahan et al.~(2023)
frame these models as ``simulators of text-generating processes''
capable of instantiating diverse ``simulacra'' (personas or roles).
Agent Identity, therefore, acts as an \textbf{Interpretability Prior}:
by casting the underlying simulator into a distinct social role, a
developer constrains behavioral unpredictability while simultaneously
giving the human operator a familiar heuristic for predicting the
agent's ``error boundaries'' and epistemic scope.

A critical boundary condition of ASAF is its explicit dependence on
individual differences in cognitive style as a formal moderating
variable. Specifically, the tendency to \emph{anthropomorphize} versus
\emph{instrumentalize} computational tools (Epley et al., 2007)
moderates the magnitude of the framework's predicted effects. ASAF
remains theoretically applicable across the entire user spectrum, but
the effect size of identity design on cognitive offloading is directly
proportional to a user's natural propensity to employ role schemata. For
users who adopt a strictly instrumentalist stance (actively resisting
role-schema activation and treating agents purely as mechanistic
software threads), the cognitive scaffolding effect is significantly
attenuated. For these instrumentalizing users, encountering role-based
design at scale (approximately 4--7 agents) may introduce cognitive
overhead rather than reduce it, as they force themselves to track raw
functional boundaries rather than relying on intuitive social
heuristics. By positioning anthropomorphic tendency as a moderating
variable rather than an absolute structural prerequisite, ASAF generates
distinct, falsifiable interaction trajectories for mixed user
populations.

\begin{center}\rule{0.5\linewidth}{0.5pt}\end{center}

\section{3. The Agentic Social Affordance Framework
(ASAF)}\label{the-agentic-social-affordance-framework-asaf}

ASAF proposes that in multi-agent AI systems with Human-in-the-Loop
configurations, agent identity design functions as a Social Affordance
layer. Before detailing its operational mechanisms, it is critical to
anchor this framework in its sociological and cognitive psychology
origins.

\subsection{3.1 Theoretical Foundations: Cognition, Sociology, and
Co-construction}\label{theoretical-foundations-cognition-sociology-and-co-construction}

In the context of ASAF, a \emph{Social Affordance} is operationally
defined as a relational property of an agent's designed identity that
signals to a human user the interaction possibilities, epistemic scope,
and appropriate collaboration modality specific to that agent's declared
role. Following Hutchby's (2001) ``third way'' between constructivist
and realist accounts of technology, ASAF treats agent identity as
neither a purely designer-imposed property (realism) nor an arbitrary
user projection (constructivism), but a relational affordance that
emerges between designed identity signals and users' pre-existing role
schemata. The relational character of social affordance entails that the
same identity label may afford different interaction possibilities
depending on the organizational, professional, and cultural context in
which it is encountered (Hutchby, 2001). ASAF's predictions are
therefore strongest for identity labels whose core cognitive posture is
cross-contextually stable---roles such as auditor, critic, or peer
reviewer carry structurally similar behavioral expectations across
cultural and institutional contexts, because their core cognitive
posture (challenging, demanding evidence, withholding acceptance) is
defined by professional function rather than local cultural convention.
Consequently, ASAF's claim that the social affordance layer constitutes
an independent design dimension (Section 1) is an \emph{analytical
framing assumption}: the social layer and the engineering layer
represent distinct classes of design considerations, drawing on distinct
design vocabularies. The framing assumption is positive rather than
definitional---it identifies a coherent set of design questions (what
cognitive posture to signal, what role schemata to activate, what
oversight to invite) that engineering decisions about orchestration,
tool exposure, and message passing do not exhaust. It is not a thesis
about effect-independence: the realized magnitude of any Social
Affordance is inherently user-moderated, scaling with the individual's
propensity to activate role schemata (see H1, Section 5), and the
layers' downstream effects can and do interact (see Section 3.4 on
persona-capability interaction). The distinction is substantive---it
concerns the structure of the design problem---but it organizes design
analysis rather than issuing a prediction about effect-independence.

A clarification on the relationship between agent identity and social
scripts is necessary. On the supply side, agents \emph{enact} culturally
established role sequences in the sense characterized by Schank and
Abelson (1977)---stereotyped action patterns with defined roles, scenes,
and behavioral expectations---drawn from professional stereotypes,
narrative media, and cultural archetypes. We use ``script'' here in a
descriptive rather than cognitive sense: the agent does not cognitively
possess a script, but its designed behavior instantiates a culturally
recognizable action sequence. On the demand side, however, users do not
necessarily require pre-existing scripts for interacting with these
enacted roles. What enables interaction initiation is social affordance
perception (Hutchby, 2001): the user recognizes the agent as the locus
where a particular cognitive function---interrogation, verification,
ideation---can be invoked. Identity Signaling (Section 3.2) serves as
the bridge between these two sides, translating the agent's enacted role
into a perceivable affordance for the human operator. The initial
affordance provides a cognitive entry point; the specific interaction
pattern is subsequently co-constructed through sustained use.

ASAF's treatment of agent identity as an ``Interpretability Prior''
inherits the instrumentalist stance on intentional description
articulated by McCarthy (1979) and operationalized for agent theory by
Wooldridge and Jennings (1995): attributing roles and social identities
to agents is legitimate not because agents possess these properties, but
because such attribution reduces the cognitive burden on human operators
managing systems whose internal structure is insufficiently transparent
for mechanistic description. ASAF extends this from a descriptive tool
for understanding individual agents to an operational interface for
collaborating with agent collectives. Therefore, the efficacy of ASAF
does not depend on the LLM possessing actual sociological awareness, but
on the human operator possessing culturally available role schemata that
agent identity design can activate. We anchor this claim in three
theoretical pillars:

First, sociological \textbf{Role Theory} (Biddle, 1986) establishes that
societal roles carry specific ``role expectations.'' ASAF does not
assume that role expectations activated by agent identity labels are
universal or automatically shared. Following Biddle's integrative
account, ASAF treats initial role expectations as structurally
primed---entering an interaction with an agent labeled ``Auditor''
activates a default behavioral script comparable to entering a hospital
and encountering a ``doctor''---but subject to ongoing negotiation
through use. Users routinely recalibrate agent identities through
continued interaction, from adjusting prompt specificity to modifying
persona configurations, a process enabled by the low technical threshold
of LLM-based systems. The initial role schema provides a cognitive entry
point; the enacted role is co-constructed through sustained
operator-agent interaction.

Second, Goffman's \textbf{Dramaturgical Analysis} (1959) introduces a
structural distinction between front stage (the performance presented to
audiences) and back stage (the private space where performances are
prepared). In LLM-based agent systems, this distinction acquires a
distinctive character: the back stage---the computational process
generating responses---is opaque to the human operator regardless of
design choices. Whether or not this process constitutes
``understanding'' is an open question (Shanahan et al., 2023); what is
not open is that the human operator cannot access it. The front
stage---the agent's identity as presented through name, role, tone, and
behavioral constraints---is therefore not a mask concealing a ``true
self'' behind it, but the operationally primary interface through which
the human operator constructs expectations, calibrates trust, and
negotiates the interaction. As in Goffman's original framework, this
front stage is not unilaterally imposed by the designer; it is
co-constructed through sustained interaction, with users continuously
recalibrating their interpretation of the agent's role.

Finally, \textbf{Distributed Cognition} (Hutchins, 1995; 2010) models
how cognitive labor is redistributed across systems comprising human,
artifactual, and representational components. Critically, Hutchins
demonstrates that distribution does not eliminate cognitive labor but
redistributes it---and the coordination required to maintain distributed
processes is itself a form of cognitive work (Hutchins, 2010). ASAF
draws on this insight with a specific qualification: identity-based role
schemata do not provide automatic cognitive routing, but they reduce the
coordination cost of navigating a multi-agent system by providing
familiar heuristic entry points for each node. The coordination labor
remains---users must still calibrate trust, verify outputs, and
negotiate role boundaries---but identity signals lower the threshold for
initiating that coordination, particularly when the number of agents
exceeds unaided working memory capacity.

These three foundational theories manifest in ASAF through three
distinct but conceptually interdependent mechanisms, operating across
three levels of analysis. Identity Signaling functions at the cognitive
level, shaping how an individual user approaches a single agent before
any interaction occurs. Behavioral Priming functions at the
interactional level, describing how that initial framing propagates into
sustained changes in user input behavior across the session.
Collaborative Governance functions at the system topology level,
operating as a bridging concept across two temporal dimensions. At
\emph{design-time}, establishing deliberate identity structures (e.g.,
configuring adversarial roles) serves as the architectural prerequisite
that enables distinct Social Affordances to exist. At \emph{run-time},
this topology governs the emergent dynamics---specifically, how human
operators differentially calibrate their oversight based on the cascaded
effects of Identity Signaling and Behavioral Priming. This structural
interplay establishes the human cognitive interface in multi-agent
systems as an independent design dimension, standing alongside the
engineering orchestration layer rather than subordinate to it.

\begin{figure}[!htbp]
\centering
\includegraphics[width=\textwidth]{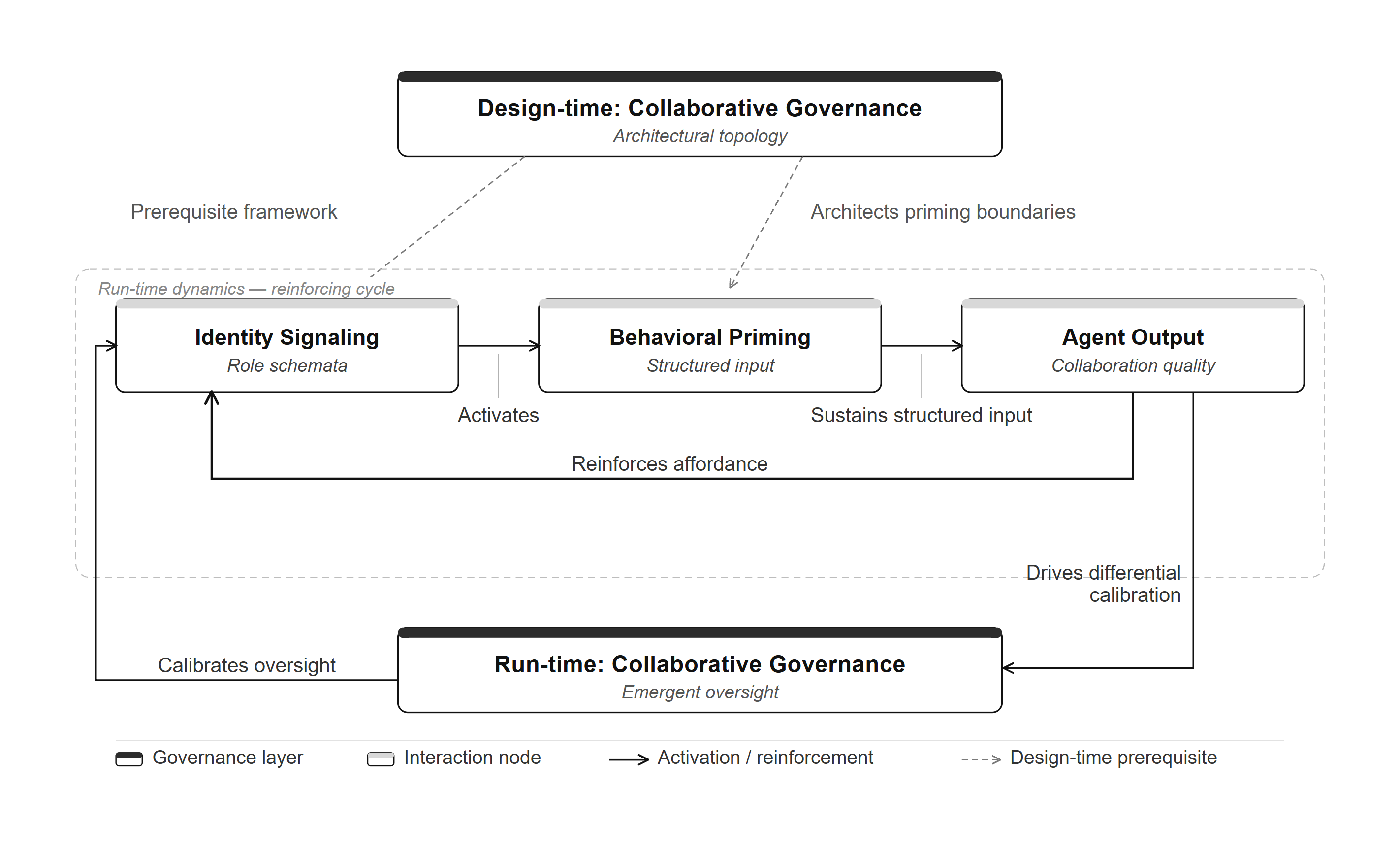}
\caption{The dual dimensions of Collaborative Governance. Design-time topology acts as the architectural prerequisite for interaction, while run-time emergent governance reflects the human operator's differential oversight. Within the run-time cycle, Identity Signaling activates role schemata and Behavioral Priming sustains structured input; Agent Output reinforces the active affordance, and drives differential calibration of oversight back into the governance layer.}
\label{fig:asaf-figure2}
\end{figure}

\subsection{3.2 Mechanism 1: Identity
Signaling}\label{mechanism-1-identity-signaling}

\textbf{Definition:} Agent identity design activates role schemata in
users before interaction begins, reducing cognitive friction and
establishing structural interaction expectations.

Recent scholarship in HCI highlights the growing necessity of
``thoughtful persona design'' in LLMs, predominantly focusing on
elements like voice, empathy, and emotional demographics to enhance
single-agent interaction quality (Zargham et al., 2024). ASAF diverges
from this paradigm. In single-agent contexts, Social Affordance operates
primarily through emotional rapport and conversational empathy---a
diffuse, system-level property with no comparative anchor. In
multi-agent contexts, this function undergoes a scale-dependent
reprioritization: as users must navigate multiple distinct agents,
structural boundary-marking between differentiated identities moves to
the foreground, while emotional rapport recedes---not stripped, but
reprioritized. This reprioritization is consistent with Hutchby's (2001)
relational account: the same technological artifact affords different
interaction possibilities depending on the complexity and structure of
the context in which it is encountered.

An Identity Signal, as used in this framework, is characterized by two
operationally distinct properties. \emph{Signal intensity} refers to the
degree to which the identity design specifies a precise cognitive
posture---the difference between ``you are an assistant'' and a fully
articulated professional archetype with defined judgment criteria,
rhetorical style, and epistemic boundaries. \emph{Signal fidelity}
refers to the degree to which the agent's enacted behavior remains
consistent with the declared identity across interactions, maintained
through persistent memory mechanisms and structural enforcement. These
two properties vary independently: a system may deploy high-intensity
signals with low fidelity (vivid personas that drift within a session)
or high-fidelity signals with low intensity (stable but vaguely defined
roles). On the supply side, signal intensity is a function of how
precisely the agent's enacted social script (Section 3.1; Schank \&
Abelson, 1977) specifies a recognizable cognitive posture; on the demand
side, it is a function of how readily the user's role schemata are
activated by that specification. The Identity Signal Fidelity Spectrum
(Section 2.3, Tiers 1--4) classifies the technical architecture by which
fidelity is maintained---the minimum structural enforcement required for
identity signals to operate reliably---rather than properties of the
signals themselves.

ASAF posits that when a user encounters an agent with a defined identity
(for instance, an architecture deploying ``Benjamin, specializing in
mathematical deduction'' alongside ``Lucas, a dedicated contrarian
analyst''), they do not begin interaction from a blank slate. We
theorize that they bring to bear culturally available role schemata for
interacting with entities who occupy similar analogous roles. These
schemata implicitly afford a sense of what kind of input is appropriate,
what level of formality is expected, and what type of output to
anticipate.

This mechanism predicts that social identity will outperform
functionally equivalent but non-personified labels (H1: condition a
\textgreater{} b) for three specific reasons. First, \emph{activation
automaticity}: role schemata are acquired through a lifetime of
interpersonal interaction, media consumption, and professional
socialization, and activate involuntarily upon encounter (Epley et al.,
2007), whereas functional descriptions (e.g., ``logical analysis
module'') require deliberate parsing to derive appropriate interaction
norms. Second, \emph{default coverage breadth}: a social role implicitly
encodes not only capability boundaries but also tone expectations,
formality level, error tolerance, and appropriate challenge
behavior---behavioral norms that a functional label does not carry.
Third, \emph{normative load}: social roles carry implicit behavioral
obligations for \emph{both parties} in the interaction (Biddle, 1986); a
user interacting with a ``critic'' feels normatively compelled to
present defensible arguments, an effect that a label reading
``verification subsystem'' does not generate. These three properties
collectively predict that social identity signals will produce richer
behavioral adaptation than equivalent functional information alone.

This pre-interaction activation is the core of Identity Signaling. The
Social Affordance operates at the moment of encounter, structuring the
user's approach to the interaction and strategically reducing the
cognitive effort required to determine ``how should I context-prompt
this system?'' This creates a specific scaling advantage in multi-agent
environments. When managing approximately 4--7 agents---the heuristic
range at which unstructured functional profiles approach working memory
limits (Cowan, 2001)---agent identities sidestep these limits by
engaging pre-existing social schemas (Epley et al., 2007) rather than
requiring the operator to manually track abstract architectural
boundaries.

This pre-interaction activation also resolves ambiguities in how users
perceive agent output parameters. Consider the counterintuitive finding
by Xu et al.~(2025), who demonstrated that expressing medium uncertainty
(e.g., ``I think this might be\ldots{}'') frequently outperforms
absolute confidence in human-AI teaming. Seen through the lens of
Identity Signaling, this makes complete sense. We do not cognitively
expect an ``ideator'' to offer design propositions with clinical
precision or formal rigor. When such an identity signals uncertainty,
the user interprets it not as a system failure, but as a culturally
legible invitation to step in and exercise human judgment.

\subsection{3.3 Mechanism 2: Behavioral
Priming}\label{mechanism-2-behavioral-priming}

\textbf{Definition:} Agent social identity is anticipated to
systematically improve the structural quality of user inputs, creating a
positive feedback loop that elevates collaboration outcomes.

Jahani et al.~(2026) document how users iteratively learn to adapt their
prompting behaviors over time, with significant performance gains
emerging from this acquired interaction fluency rather than from model
improvements alone.

ASAF's Behavioral Priming mechanism theorizes how agent identity drives
this influence. We postulate that when interacting with what they
tacitly treat as a domain-competent counterpart (cf.~Reeves \& Nass,
1996), users are primed to voluntarily: * Provide more structured,
context-rich inputs aligned directly with the agent's declared domain. *
Proactively include constraints, assumptions, and boundary conditions. *
Formulate questions at a higher level of structural specificity.

Jahani et al.~(2026) document that this fluency typically develops
through iterative trial-and-error across multiple sessions---a process
they characterize as a \emph{dynamic complement}, noting that adaptation
imposes real costs on users and that failure to adapt degrades outcomes.
ASAF hypothesizes that strong Identity Signaling may reduce the initial
adaptation cost by providing role-appropriate cognitive entry points
from the first interaction, though the extent of this compression
remains an empirical question.

A secondary, yet crucial implication of Behavioral Priming relates to
cognitive bias mitigation. It is well-documented that human users are
dangerously susceptible to AI overconfidence, often failing to detect
when a model is hallucinating assertively (Li et al., 2024). However,
what happens if we deliberately assign the system a socially cautious or
adversarial persona? If a user explicitly knows they are interacting
with an ``auditor'' (by definition, a role whose enacted behavior
foregrounds challenge and verification over generative output), this
alone structurally intercepts their default vulnerability to
over-reliance. The social frame does the work; no explicit instruction
to ``be skeptical'' is required. We further note that in LLM-based
systems where designer and operator are coupled, Behavioral Priming
co-evolves with operator learning---a structural condition we treat as a
feature of LLM agent research methodology rather than a confound to be
controlled away (elaborated in future work).

\subsection{3.4 Mechanism 3: Collaborative
Governance}\label{mechanism-3-collaborative-governance}

\textbf{Definition:} In Human-in-the-Loop configurations, agent social
identity structures the location and mode of human intervention,
transforming HitL from a mere safety checkpoint into a collaboration
design element.

The conventional framing of Human-in-the-Loop in AI systems is
safety-centric: human oversight exists purely to catch errors. This
framing is correct but incomplete. The foundational literature on
automation reliance (Parasuraman \& Riley, 1997) and the dynamics of
human trust in automated systems (Lee \& See, 2004) establishes that
human operators constantly modulate their reliance and oversight based
on the system's perceived purpose and past performance. ASAF extends
this to the multi-agent context, positing that when a system possesses
defined agent identities, users do not intervene generically; their
oversight calibration is dynamically structured by each agent's specific
social role. This governance operates across two temporal scopes. At
\emph{design-time}, the framework prescribes verifying that newly
introduced agents activate distinct role schemata rather than providing
redundant Social Affordance signals. At \emph{run-time}, it prescribes
monitoring for epistemic convergence between agents whose identities
were intended to remain distinct. Section 4.2 operationalizes both
functions in the ChronicleCore implementation. In a recent scoping
review of 134 top-tier HCI papers, Zhang et al.~(2025) constructed an
integrated theoretical map of how ``agency'' and control are distributed
in human-AI co-creation. ASAF's Collaborative Governance mechanism
serves to operationalize the specific control dynamics mapped by Zhang.
We argue that assigning a social identity to an agent is a specific
method of \emph{agency negotiation}. A user understanding that one agent
is an ``ideator'' and another is an ``auditor'' will trust the ideator's
divergence while heavily scrutinizing the auditor's final checks,
thereby calibrating how deeply to intervene at each stage of the
workflow.

This mechanism extends into group dynamics. Empirical research validates
that groups of AI agents can function as social groups, exerting
``normative social pressure'' on human users, particularly through
conformity effects when multiple agents agree (Song et al., 2025). Under
ASAF's Collaborative Governance, this pressure can be architecturally
countered. By deliberately designing interacting agents with competing
identities (e.g., assigning a ``Devil's Advocate'' identity to
explicitly challenge a ``Proposer'' identity), developers can
artificially break the unanimity that causes dangerous conformity
effects, structurally forcing the human operator to exercise critical
judgment. The governance structure of the system becomes legible and
manipulable through the Social Affordance layer. What was once invisible
coordination logic is now a perceivable, interactable interface that the
human operator can navigate directly.

A critical in-scope boundary condition of Collaborative Governance is
the \textbf{false-specialization inversion}. When Identity Signaling
produces false confidence in agent specialization that the underlying
model cannot deliver---for example, an agent titled ``Data Verification
Analyst'' that hallucinates frequently---the social affordance
\emph{inverts}. Under ASAF's own governance predictions, this inversion
produces worse trust miscalibration than if no identity had been
assigned at all, because the user's oversight has been calibrated to a
social contract the agent cannot honor. This false-specialization
inversion addresses Yan's (2025) critique of multi-agent fragility from
within ASAF's framework: Yan's concern that artificial role boundaries
induce reasoning drift is partly cognitive---it concerns the
misalignment between operator mental models, primed by role assignment,
and the model's actual capability surface. ASAF therefore predicts both
when identity design helps and when it actively harms. In LLM-based
systems, persona specification and behavioral capability exhibit
interaction effects---persona instructions can shift the distribution
from which agent responses are sampled. The framework treats analytical
separability as an organizing assumption about \emph{design
vocabularies}; the design vocabularies remain distinct (configuring
``what cognitive posture to signal'' is a different question from
configuring ``what tools to expose'') even though the resulting system
properties---including how persona instructions reshape the sampling
distribution---are jointly determined.

Taken together, the preceding mechanisms make explicit that ASAF
predicts three categories of anthropomorphism-related risk that are
integral to the framework rather than peripheral concerns. First,
\emph{cognitive overhead risk}: for users with strong instrumentalizing
tendencies, role-based design at scale (approximately 4--7 agents)
imposes interpretive cost rather than reducing it (Section 2.4). Second,
\emph{trust miscalibration risk}: the false-specialization inversion
formalized above predicts that misleading identity signals produce worse
oversight outcomes than the absence of identity assignment, because
operator oversight has been calibrated to a social contract the agent
cannot honor. Third, \emph{overconfidence-attenuation failure}: ASAF
positions adversarial or auditor-typed identity signals as a structural
intervention against user susceptibility to AI overconfidence (Section
3.3), with the corollary that omitting such identity scaffolding leaves
operators exposed to default overreliance patterns documented in the
human-AI trust literature (Lee \& See, 2004; Li et al., 2024). These
three categories are core ASAF predictions, not externalities to be
deferred. The boundary phenomena discussed in Section 7 (parasocial
dynamics, cross-cultural variation, dynamic identity formation)
represent unmeasured outer surfaces of these core risks rather than
substitutes for them, and constitute the priority directions for
empirical extension.

\begin{center}\rule{0.5\linewidth}{0.5pt}\end{center}

\section{4. System Implementation: The Design Problem That Motivated
ASAF}\label{system-implementation-the-design-problem-that-motivated-asaf}

Rather than presenting empirical validation, we offer the development of
\emph{ChronicleCore}---an evolving, single-operator Multi-Agent
System---as \textbf{the design problem that motivated ASAF}: a case that
illustrates how the framework's design space might be navigated in
practice, and that surfaces the questions ASAF aims to theorize.
Validation of the framework's predictions is explicitly deferred to the
empirical designs specified in H1--H3 (Section 5). The researcher
simultaneously occupied the roles of system designer, primary operator,
and observing analyst. This configuration is useful for generating
design rationale and surfacing tacit operational knowledge, but it does
not constitute independent empirical validation---a limitation the
author acknowledges without reservation.

This paper was itself produced under a multi-agent AI collaboration
configuration (see Acknowledgments for specific tools and versions). All
conceptual contributions (the ASAF framework, its mechanisms, and
hypotheses) originated entirely from the human author and all cited
works were manually verified prior to inclusion. The author acknowledges
the circular evidentiary structure this creates---using a multi-agent
configuration to argue for the value of multi-agent configurations. This
methodological reflexivity is part of the framework's boundary
conditions rather than its validation.

At the time of writing (April 2026), ChronicleCore comprises 38 active
Human-in-the-Loop expert nodes, designed to organically accommodate
incremental agent addition as operational needs dictate. Operated by the
author as a personal productivity architecture, this case illustrates
the design challenges that arise when a single operator must navigate a
system whose scale exceeds unaided working memory. Unlike multi-agent
simulations of human societal behavior (Park et al., 2023),
ChronicleCore utilizes deliberately designed societal structures (agent
identities) to structure the reasoning pathways of a single human
operator.

\subsection{4.1 Decoupling Agents by Epistemic
Role}\label{decoupling-agents-by-epistemic-role}

When a single operator scales a system to several dozen specialized
agents, the primary challenge transcends tool orchestration and enters
the realm of personal cognitive governance. A single, monolithic chatbot
affords generic conversational prompting, but a highly populated agent
network risks cognitive overload for the human operator and ``persona
drift'' among the agents themselves. To counteract this, ChronicleCore
physically decouples the system into distinct operational pillars, each
functioning as a highly specific Social Affordance boundary.

Rather than deploying agents as indistinguishable utility nodes, the
architecture categorizes them by their epistemic and rhetorical
constraints. The system separates strategic routing nodes (which
allocate global context but possess no execution capabilities) from
sensory nodes responsible purely for data ingestion. It isolates
aesthetic alignment tasks from adversarial verification layers. By
granting each agent localized context paired with a heavily defined
social identity (e.g., naming a verification node the
``Inquisitor''---see Appendix A for a summary of the agent identity
definition and Appendix B, §B.3 for the full Iron Laws
table), the system is designed to structurally force the human operator
to calibrate their oversight. This architectural decoupling was designed
precisely to shape the operator's interaction patterns. By structurally
enforcing the separation---where an adversarial node is configured
(through prompt instruction and capability gating) to focus exclusively
on probing logical vulnerabilities rather than generative code
production---the system aims to preempt persona drift (subject to the
persona-capability interaction effects acknowledged in Section 3.4). The
design enforces the boundary, ensuring that the operator cannot bypass
the intended Social Affordance without breaking the system's operational
protocol.

\subsection{4.2 Sustaining Identity Boundaries: A Design
Perspective}\label{sustaining-identity-boundaries-a-design-perspective}

The persistence challenge in a multi-agent system with differentiated
social identities is not only intra-agent but inter-agent: ensuring that
Agent A's epistemic and rhetorical profile does not assimilate toward
Agent B's over repeated interactions. ChronicleCore's design addresses
this through two operational practices.

\emph{Memory Crystallization} separates transient reasoning logs from
identity-weighted constraint rules encoded in each agent's persistent
SKILL.md layer; only the latter are promoted into long-term state,
shielding the agent's social identity from accumulated task-specific
reasoning drift. \emph{Personality Variance Audit} is a systemic
cross-agent inspection---operationalized in ChronicleCore through the
Inquisitor node's VETO threshold (Iron Law 5; see Appendix A for a
summary and Appendix B, §B.3 for the full Iron Laws
table)---that detects rhetorical and epistemic convergence between
agents intended to maintain distinct social identities. Whereas the
Personality Uniqueness Check (the design-time distinctness-check
principle articulated in §6) operates at design-time to prevent
redundant social affordance signals when new agents are introduced, the
Personality Variance Audit operates continuously at runtime to detect
epistemic convergence between already-deployed agents.

These are design patterns for applying existing memory mechanisms in
service of Social Affordance integrity. The design rationale behind
these patterns instantiates the question ASAF aims to theorize: whether
sharply defined, non-overlapping identities facilitate distinct human
interaction patterns, while blurred identities give way to operational
genericism. Whether this design rationale produces the predicted effects
is an empirical question deferred to H1--H3.

\begin{center}\rule{0.5\linewidth}{0.5pt}\end{center}

\section{5. Formalizing ASAF: Testable
Hypotheses}\label{formalizing-asaf-testable-hypotheses}

To transition ASAF from a descriptive framework into an empirically
verifiable model, we formulate three core hypotheses aligned with
Human-Computer Interaction (HCI) methodologies. These hypotheses
explicitly address potential confounds, such as individual baseline
tendencies (anthropomorphizing vs.~instrumentalizing) and alternative
explanations (the novelty effect), ensuring the framework's
falsifiability:

\begin{itemize}
\tightlist
\item
  \textbf{H1 (Identity Signaling and Interaction Framing):} Users
  interacting with agents bearing differentiated role identities will
  produce first-turn prompts exhibiting greater domain alignment than
  users interacting with generic labels. To discriminate between the
  effects of social identity and mere task decomposition clarity,
  validation should employ a three-condition design: (a)
  \emph{identity-differentiated} agents (persona, name, tone), (b)
  \emph{functionally labeled} agents (capability descriptions without
  personification), and (c) \emph{generic} agents (no label). ASAF
  predicts condition (a) \textgreater{} (b) \textgreater{} (c); if the
  effect derives solely from task decomposition clarity, conditions (a)
  and (b) should not differ significantly. This contrast tests the
  independent explanatory power of social identity design beyond
  functional labeling; characterizing the broader interaction surface
  between the social and engineering layers would require a factorial
  design that simultaneously manipulates both (see Section 7).
  Furthermore, at or beyond the multi-agent scaling heuristic
  (approximately 4--7 agents), this effect is significantly moderated by
  baseline tendencies: users with high anthropomorphizing tendencies
  will leverage social identities to route tasks with significantly
  lower perceived cognitive load, whereas strict instrumentalizing users
  will experience rapid cognitive degradation as they attempt to
  manually track unstructured parameters. \emph{Operationalization.} The
  moderating variable (anthropomorphizing tendency) should be measured
  via the Individual Differences in Anthropomorphism Questionnaire
  (IDAQ; Waytz, Cacioppo, \& Epley, 2010), which provides a validated
  trait-level instrument with established psychometric properties.
  Domain alignment of first-turn prompts can be quantified through
  domain-specific lexical density (using a pre-registered domain
  lexicon) cross-validated against expert relevance ratings (target
  inter-rater reliability κ ≥ 0.7). To isolate personification as the
  manipulated variable, information quantity must be held constant
  across conditions (a) and (b) by matching word count and propositional
  content between identity labels and functional descriptions.
  \emph{Falsification.} H1 is rejected if the predicted (a)
  \textgreater{} (b) contrast fails to reach significance at adequate
  statistical power (anticipated effect size: Cohen's \emph{d} ≥ 0.3).
\item
  \textbf{H2 (Behavioral Priming vs.~Novelty Effect):} Over a sustained
  task session, the proportion of structured inputs (e.g., constraint
  density, explicit boundary conditions) will be significantly higher in
  identity-differentiated conditions than in generic-label conditions.
  H2's validation design is a between-subjects experiment in which
  participants are randomly assigned to identity-differentiated,
  functionally labeled, or generic conditions without the ability to
  modify the agent's identity configuration, thereby isolating identity
  design effects from user-driven adaptation. To distinguish this from a
  mere \textbf{novelty effect}, these behavioral adaptations will not
  decay over time. Instead, adherence to the agent's role-based
  interaction pattern will exhibit a dose-response relationship with the
  intensity of the identity signal (e.g., consistency of tone, boundary
  reinforcement) and will stabilize over repeated longitudinal
  interactions. \emph{Falsification.} H2 is rejected if structured-input
  rates in identity-differentiated conditions converge to
  generic-condition rates within five sustained sessions, or if the
  predicted dose-response relationship between signal intensity and
  structured-input proportion fails to reach significance (anticipated
  effect size: Cohen's \emph{d} ≥ 0.4 for the strongest intensity
  contrast).
\item
  \textbf{H3a (Collaborative Governance --- Role-Level):} In
  Human-in-the-Loop multi-agent architectures, human intervention
  patterns will significantly differ across agents bearing different
  role identities. Specifically, users will exhibit higher unverified
  acceptance of generative outputs from designated ``Ideator''
  identities, while applying significantly higher scrutiny to outputs
  from assigned ``Verifier/Auditor'' identities, demonstrating
  deliberate oversight calibration. This prediction can also be
  generated by CASA-plus-role-schema accounts (Gambino et al., 2020;
  Biddle, 1986) and therefore does not, by itself, distinguish ASAF from
  existing frameworks. \emph{Falsification.} H3a is rejected if
  oversight metrics (acceptance latency, edit rate, explicit
  verification queries) do not differ significantly between Ideator and
  Auditor agents under role-differentiated conditions (anticipated
  effect size: Cohen's \emph{d} ≥ 0.3).
\item
  \textbf{H3b (Collaborative Governance --- Topology-Level):} The same
  agent identity will elicit different oversight behavior depending on
  its topological context within the multi-agent system. Specifically,
  an ``Auditor'' embedded in a surround occupying cognitive postures
  that contrast with the focal Auditor's evaluative function (a
  \emph{competing-posture} surround comprising, for instance,
  generative, counter-argumentative, and integrative roles) should
  produce different operator oversight than the same ``Auditor''
  embedded in a surround occupying postures aligned with the focal
  Auditor's evaluative function (a \emph{consensus-posture} surround
  comprising, for instance, multiple complementary verification roles).
  H3b is, in the framework's reading, the prediction that
  CASA-plus-role-schema accounts cannot generate, because it requires
  sensitivity to the relational configuration of surrounding agents---a
  systemic property that dyadic frameworks do not model. If H3b is
  empirically supported while H3a alone does not discriminate ASAF from
  CASA, H3b constitutes ASAF's clearest empirical signature.
  \emph{Falsification.} H3b is rejected if the oversight metrics
  specified for H3a, when measured for the focal Auditor identity,
  remain statistically equivalent across topological conditions
  (competing-posture surround vs.~consensus-posture surround); the
  predicted interaction effect of agent identity × topological context
  should reach ηp² ≥ 0.05. A worked validation design is given
  immediately below.
\end{itemize}

\emph{Worked Validation Design for H3b.} Because H3b is, in the
framework's reading, ASAF's clearest empirical signature, its validation
plan requires sufficient specification to demonstrate that the predicted
identity × topology interaction can in principle be isolated from
confounds in surround behavior. H1, H2, and H3a admit standard
experimental designs whose validation pathways are summarized in Section
7. H3b warrants the additional structural specification given below
because the topology-level interaction it predicts requires explicit
isolation from alternative mechanisms operating in dyadic frameworks
(see Section 1 for the full demarcation from CASA and
CASA-plus-role-schema accounts). The following is an \emph{illustrative}
worked design, not a finalized protocol; it makes the manipulation logic
and confound controls concrete enough to be auditable, while leaving
role instantiation, sample size, and equivalence thresholds to be
calibrated by a preceding pilot study.

\emph{Condition structure (between-subjects, single-factor topology
manipulation).} The manipulation contrasts two surround topologies for a
focal Auditor whose identity is held constant. To bring the
configuration within the multi-agent scaling threshold identified in
§2.4 (approximately 4--7 agents) and to match cognitive-role diversity
across conditions (so that the manipulation isolates surround
\emph{posture} rather than surround \emph{diversity}), each condition
uses a four-agent configuration with three surround roles:

{\def\LTcaptype{none} 
\begin{longtable}[]{@{}
  >{\raggedright\arraybackslash}p{(\linewidth - 4\tabcolsep) * \real{0.3333}}
  >{\raggedright\arraybackslash}p{(\linewidth - 4\tabcolsep) * \real{0.3333}}
  >{\raggedright\arraybackslash}p{(\linewidth - 4\tabcolsep) * \real{0.3333}}@{}}
\toprule\noalign{}
\begin{minipage}[b]{\linewidth}\raggedright
Condition
\end{minipage} & \begin{minipage}[b]{\linewidth}\raggedright
Focal agent (held constant)
\end{minipage} & \begin{minipage}[b]{\linewidth}\raggedright
Surround posture configuration (illustrative)
\end{minipage} \\
\midrule\noalign{}
\endhead
\bottomrule\noalign{}
\endlastfoot
Competing-posture surround & Auditor (identity Φ) & Ideator (generative)
· Devil's Advocate (counter-argumentative) · Synthesizer
(integrative) \\
Consensus-posture surround & Auditor (identity Φ) & Evidence-Verifier ·
Logic-Verifier · Source-Verifier \\
\end{longtable}
}

The instantiation in this table is illustrative. In both conditions the
surround comprises three distinct cognitive roles, so that \emph{role
diversity} (number of distinct cognitive postures present in the
surround) is held constant. What varies across conditions is the
\emph{posture relation} between the surround and the focal
Auditor---whether the surround occupies postures that contrast with the
focal Auditor's evaluative function (competing condition) or align with
it (consensus condition). The focal Auditor's identity label, role
description, tone, output content, and turn timing are bit-identical
across conditions. To isolate operator response to topology from
operator response to LLM-generated variability, the focal Auditor's
outputs are pre-scripted (Wizard-of-Oz protocol); see \emph{Scope of
Generalization} below for the trade-off this introduces.

\emph{Confound controls.} Four classes of surround-behavior confound are
addressed. (i) \emph{Output quality} is matched by configuring all
surround agents to emit outputs of equivalent length (within a
pre-registered tolerance band), equivalent evidence-citation density,
and structurally parallel argument schemata. A pre-study using expert
raters verifies that surround agents in both conditions are perceived as
equivalently competent and equivalently authoritative; the equivalence
criterion is formalized via a Two One-Sided Tests (TOST) procedure
rather than a non-significance threshold, with the equivalence bound
established by the pilot study. (ii) \emph{Salience} is matched by
fixing turn order across conditions, holding interface representation
(avatar size, font, message styling) identical across all surround
agents, and matching the word count of role labels and role descriptions
so that the descriptive footprint of each surround role is comparable.
(iii) \emph{Turn frequency} is matched by allocating each surround agent
the same number of turns in both conditions, and by constraining total
system output (words per session) to within a pre-registered tolerance
band between conditions. (iv) \emph{Cognitive-role diversity} is matched
by ensuring that both conditions present three distinct surround
cognitive roles, so that the manipulation isolates the relational
posture of the surround rather than the count of distinct roles present.

\emph{Operator-oversight metrics (the dependent variables).} Measurement
targets the focal Auditor only, not the surround. The four metrics,
drawn from H3a's instrumentation, are: acceptance latency (seconds from
focal output to operator decision), edit rate (proportion of focal
outputs the operator modifies before accepting), verification-query
count (explicit clarification or evidence-request follow-ups directed at
the focal Auditor), and override rate (proportion of focal claims the
operator countermands in the final synthesis). The scoring rubric is
identical across conditions; only the surround topology varies.

\emph{Isolation logic.} Because the focal Auditor's outputs are
bit-identical, any condition-level difference in oversight of the focal
Auditor cannot be attributed to changes in the Auditor's behavior.
Because surround output quality, salience, turn frequency, and
cognitive-role diversity are pre-equated, any condition difference
cannot be attributed to one surround being ``louder,''
``better-argued,'' ``more frequent,'' or ``more diverse'' than the
other. The residual variance is attributable to relational posture
configuration---whether the surround occupies postures contrasting with
the focal Auditor (competing surround) or aligned with it (consensus
surround). ASAF predicts reduced oversight of an Auditor in a consensus
surround (relational redundancy reduces the operator's perceived need
for independent verification) and elevated oversight in a competing
surround (relational distinctness renders the Auditor's epistemic
function more salient).

\emph{Pre-registration commitments.} The surround-equivalence pre-study
runs before the confirmatory experiment and calibrates (a) the
equivalence bound for the TOST procedure, (b) the manipulation-check
thresholds for perceived posture-distance, and (c) the target sample
size derived from the pilot effect size estimate. The confirmatory study
is registered before data collection with the pre-specified analysis
plan and exclusion rules; sessions failing the manipulation check are
excluded from confirmatory analysis but reported in a sensitivity
analysis.

\emph{Scope of generalization.} This worked design buys variable
isolation at the cost of two ecological-validity trade-offs that should
be explicit. First, Wizard-of-Oz scripting of the focal Auditor controls
for LLM output variability, but the resulting study tests operator
response to a \emph{controlled relational configuration} rather than to
emergent multi-agent dynamics. Second, the four-agent configuration sits
at the lower bound of the scaling threshold identified in §2.4; the
topology manipulation may exhibit different magnitudes---or qualitative
shifts---at higher agent counts. Generalization to fully autonomous
LLM-driven multi-agent systems and to larger-scale topologies (8+
agents) requires follow-up studies that progressively release these
controls. The present design is therefore best read as the \emph{first}
validation step in a programme: an initial isolation of the predicted
interaction, followed by ecological-validity-recovering extensions.

\emph{Cross-Cultural Role-Label Norming.} Consistent with ASAF's
acknowledgment that role schemata are not culturally universal (Section
3.1), H1--H3b's validation plan incorporates a preliminary norming step
prior to hypothesis testing. Candidate identity labels (e.g.,
``auditor,'' ``critic,'' ``reviewer,'' ``inquisitor'') should be
assessed for perceived cognitive posture, authority, threat valence, and
collaboration expectation across the relevant user populations sampled.
Where a label exhibits significant cross-cultural or cross-institutional
variation in perceived posture, the corresponding hypotheses should be
tested within rather than across population strata, and results should
be reported with explicit scope conditions tied to the validated label
set. Comprehensive cross-cultural validation of the framework's
predictive scope is deferred to follow-up empirical work; the norming
step specified here is intended to ensure that within-study label
interpretation is internally consistent rather than to settle the
broader question of cross-cultural portability.

\begin{center}\rule{0.5\linewidth}{0.5pt}\end{center}

\section{6. Implications for Multi-Agent System
Design}\label{implications-for-multi-agent-system-design}

ASAF suggests several design principles for practitioners building
multi-agent systems with human interaction layers:

\begin{itemize}
\tightlist
\item
  \textbf{Design agent identity for the interaction, not the
  implementation.} The social identity of an agent should be designed
  from the perspective of what interaction pattern it should elicit from
  users, not from the perspective of what the underlying model can do.
  Implementation capabilities constrain the design space; Social
  Affordance considerations should guide the design choices within that
  space.
\item
  \textbf{Differentiate identities meaningfully.} If two agents in a
  system have social identities that elicit similar interaction patterns
  from users, they are functionally equivalent from a Social Affordance
  perspective---regardless of their technical differences. Meaningful
  differentiation requires that each agent's identity activates a
  distinct role schema; agents whose behavioral and epistemic profiles
  are insufficiently distinct provide redundant Social Affordance
  signals. Section 4.2 describes how ChronicleCore operationalizes this
  design-time distinctness check.
\item
  \textbf{Treat HitL configuration as Social Affordance design.} The
  placement and nature of human intervention points in a multi-agent
  system should be designed in relation to agent identities, not
  independently of them. Human oversight is most effective when users
  understand, through Social Affordance signals, what each agent is
  responsible for and what kind of human judgment is most relevant at
  each intervention point.
\item
  \textbf{Separate the social layer from the orchestration layer.} Agent
  identity design (the social layer) and agent orchestration design (the
  engineering layer) address different problems and should be designed
  with different considerations. Conflating them---designing agent roles
  purely based on engineering decomposition---risks producing social
  identities that are technically coherent but interaction-incoherent.
\end{itemize}

\begin{center}\rule{0.5\linewidth}{0.5pt}\end{center}

\section{7. Limitations and Future
Research}\label{limitations-and-future-research}

While ASAF is theoretically grounded and illustrated through the
ChronicleCore design case, its mechanisms have not yet been subjected to
controlled, large-scale empirical validation. Several limitations follow
from this:

\begin{itemize}
\tightlist
\item
  \textbf{Measurement challenges.} Social Affordance effects are
  inherently difficult to isolate and quantify. The influence of agent
  identity on user behavior is confounded by user expertise, task
  complexity, system context, and individual differences in role-schema
  activation. Future research should develop measurement instruments
  adapted from existing Social Affordance and Human-AI interaction
  scales, with individual-difference moderation linked to established
  instruments such as the IDAQ (Waytz et al., 2010).
\item
  \textbf{Context dependency and cultural variation.} ASAF's mechanisms
  are likely to vary in salience across different deployment contexts.
  Enterprise users with high AI literacy may exhibit different Social
  Affordance responses than general consumers. Task domains with strong
  professional role norms (legal, medical, financial) may show stronger
  Identity Signaling effects than open-ended creative tasks. More
  fundamentally, the role schemata that agent identity labels activate
  are not culturally universal (Biddle, 1986; Hutchby, 2001). ASAF's
  predictions are strongest for identity labels drawing on professional
  role norms with cross-cultural stability---roles such as auditor,
  reviewer, or analyst whose core cognitive posture is defined by
  professional function rather than local cultural convention. Identity
  labels with culturally specific connotations may produce variable
  affordance effects, and cross-cultural validation of such labels
  constitutes a priority for future empirical work.
\item
  \textbf{The anthropomorphism boundary.} ASAF deliberately engages with
  anthropomorphic design elements in agent identity. The literature on
  human-AI interaction has documented both the benefits and risks of
  anthropomorphism, including parasocial relationship formation and the
  potential for contextually inappropriate overtrust in faulty systems
  (Hancock et al., 2011; Salem et al., 2015). The false-specialization
  inversion formalized in Section 3.4 addresses one dimension of this
  risk; future work should map the full boundary between beneficial and
  harmful anthropomorphism across deployment contexts.
\item
  \textbf{Misleading identity signals.} The false-specialization
  inversion (Section 3.4) establishes that misleading identity signals
  can produce worse outcomes than no identity assignment at all.
  Formalizing the degradation curve of misleading identity signals---how
  rapidly trust miscalibration escalates as the gap between declared and
  enacted identity widens---represents a crucial priority for future
  study.
\item
  \textbf{Dynamic identity formation.} ASAF's current treatment of
  identity signals is predominantly static: it theorizes how designed
  signals affect user behavior at the point of encounter. However,
  operator-agent interaction is inherently dynamic---users recalibrate
  their interpretation of agent identity through sustained use, and the
  specific interaction patterns that emerge are co-constructed rather
  than designer-determined (cf.~Biddle, 1986; Goffman, 1959). The
  process by which system-specific interaction patterns dynamically form
  through repeated operator-agent interaction constitutes a distinct
  phenomenon that ASAF identifies but does not fully theorize in this
  paper.
\end{itemize}

Additionally, although ASAF treats the analytical separability of the
social and engineering design layers as a framing assumption (Section
3.1) rather than as a testable claim about effect-independence, the
empirical interaction surface between the two layers warrants direct
characterization. A 2×2 factorial design crossing social identity
intensity (high vs.~low personification) with engineering orchestration
quality (robust vs.~fragile architecture) would map how the layers
interact at the effect level. The goal is not to validate orthogonality,
which is an analytical commitment rather than an empirical prediction.
The goal is to understand under what engineering conditions
identity-design effects emerge, attenuate, or invert. The framework does
not commit to a specific interaction pattern; the factorial design
serves to chart the empirical surface that the framing assumption
brackets.

This framework deliberately focuses on how social affordance structures
human operators' cognition and oversight. The directional
complement---how agents might calibrate their mutual interactions
through formalized norms and role structures---is a mechanistically
distinct problem that falls within the normative MAS tradition (Dignum,
2004; Stamper, 1996; Wooldridge \& Jennings, 1995). Future work may
explore the integration of these two directions into a dual-facing
design space: human-facing affordances governing oversight, agent-facing
norms governing coordination.

Each of the ASAF hypotheses admits distinct validation pathways for
future research: 1. \textbf{H1 (Identity Signaling)} is most directly
tested through controlled studies comparing first-turn user input
quality and domain-alignment across systems with well-defined versus
generic agent identities, analyzed using NLP metrics for domain-specific
lexical density or expert-rated relevance scores. 2. \textbf{H2
(Behavioral Priming)} requires longitudinal between-subjects designs
tracking how Social Affordance effects evolve as users gain familiarity
with a multi-agent system, quantifying structural elements (e.g.,
constraint density, explicit boundary conditions) over sustained task
sessions. 3. \textbf{H3a (Role-Level Governance)} is testable through
standard between-subject experiments comparing intervention patterns
across role-differentiated versus generic agent configurations. 4.
\textbf{H3b (Topology-Level Governance)} requires experimental
manipulation of the topological configuration surrounding a target agent
while holding the target agent's identity constant, testing whether the
same identity elicits different operator oversight as a function of
surrounding agent roles.

\begin{center}\rule{0.5\linewidth}{0.5pt}\end{center}

\section{8. Conclusion}\label{conclusion}

The design of multi-agent AI systems has been dominated by engineering
concerns. This paper argues that as these systems are deployed in
Human-in-the-Loop configurations, a parallel design challenge emerges:
how the social identity of individual agents shapes human behavior and
collaboration quality.

The Agentic Social Affordance Framework (ASAF) provides a conceptual
vocabulary for addressing this challenge. By identifying Identity
Signaling, Behavioral Priming, and Collaborative Governance as the core
mechanisms through which agent identity design operates, ASAF offers a
foundation for both design practice and future empirical research.

The central claim of ASAF is that as multi-agent systems scale beyond
human working memory capacity, agent identity design becomes a
structural collaboration interface, not a UX convention. Deliberate
Social Affordance design establishes the human cognitive interface as an
independent design dimension, one that scales with system complexity and
becomes increasingly consequential as multi-agent architectures grow
beyond the boundaries of unaided human working memory.

\begin{center}\rule{0.5\linewidth}{0.5pt}\end{center}

\section{Data Availability Statement}\label{data-availability-statement}

The original contributions presented in the study are included in the
article/Supplementary Material. The ChronicleCore system architecture
referenced in this paper is documented in a public code repository (Lee,
2026; https://github.com/Zaious/ChronicleCore-Architecture). Further
inquiries can be directed to the corresponding author.

\section{Author Contributions}\label{author-contributions}

M-HL: Conceptualization, Methodology, Writing -- original draft, Writing
-- review \& editing.

\section{Funding}\label{funding}

This research received no external funding.

\section{Conflict of Interest}\label{conflict-of-interest}

The author declares that the research was conducted in the absence of
any commercial or financial relationships that could be construed as a
potential conflict of interest.

\section{Acknowledgments}\label{acknowledgments}

I am grateful to the Department of Interaction Design at National Taipei
University of Technology, my alma mater, whose grounding in design
psychology and design thinking, together with my academic training in
user experience design, shaped the intellectual foundation for this
work. I thank the handling editor for taking on this Hypothesis and
Theory submission---my first as an independent researcher---and for
identifying reviewers who could engage with the conceptual context of
the proposed framework. The reviewers' expert, specific suggestions
surfaced gaps in the theoretical argumentation and substantially
strengthened this manuscript, for which I am grateful. I also thank two
fellow researchers who reviewed early drafts before the initial arXiv
submission attempt: one for pointing out a critical gap in the paper's
framing and positioning, and the other for providing the first
affirmation of the topic through an arXiv endorsement.

\textbf{Generative AI Disclosure.} The following AI tools were used
during manuscript preparation: Gemini 3.1 Pro (Google DeepMind) was used
to surface candidate reference literature and assist with drafting;
Claude Opus 4.7 (Anthropic) was used to flag logical inconsistencies
during the review of theoretical foundations and to facilitate
methodological formatting; Claude Sonnet 4.6 (Anthropic) was
additionally used to generate the HTML/CSS source code for the academic
line-art styling of Figures 1 and 2, with the figure structure,
conceptual content, and final visual review performed by the human
author. All conceptual contributions---including the ASAF framework, its
three mechanisms, and all hypotheses---originated entirely from the
human author. All cited works were manually verified by the author prior
to inclusion. The AI tools were not used to generate the theoretical
arguments, interpret data, or draw conclusions presented in this paper.
The author assumes full responsibility for the accuracy and integrity of
all content.

\begin{center}\rule{0.5\linewidth}{0.5pt}\end{center}

\subsection{References}\label{references}

\begin{itemize}
\tightlist
\item
  Amershi, S., Weld, D., Vorvoreanu, M., Fourney, A., Nushi, B.,
  Collisson, P., Suh, J., Iqbal, S., Bennett, P. N., \& Inkpen, K.
  (2019). Guidelines for Human-AI Interaction. In \emph{Proceedings of
  the 2019 CHI Conference on Human Factors in Computing Systems} (CHI
  '19), Article 3, 1--13. ACM. https://doi.org/10.1145/3290605.3300233
\item
  Anthropic. (2025a, September). Effective context engineering for AI
  agents. \emph{Anthropic Engineering Blog}. Retrieved April 19, 2026,
  from
  https://www.anthropic.com/engineering/effective-context-engineering-for-ai-agents
\item
  Anthropic. (2025b). Skills. \emph{Claude Code documentation}.
  Retrieved April 19, 2026, from https://code.claude.com/docs/en/skills
\item
  Bansal, G., Nushi, B., Kamar, E., Lasecki, W. S., Weld, D. S., \&
  Horvitz, E. (2019). Beyond accuracy: The role of mental models in
  human-AI team performance. \emph{Proceedings of the AAAI Conference on
  Human Computation and Crowdsourcing (HCOMP)}, 7, 2-11.
\item
  Biddle, B. J. (1986). Recent developments in role theory. \emph{Annual
  Review of Sociology}, 12(1), 67-92.
\item
  Boyd, d.~(2010). Social network sites as networked publics:
  Affordances, dynamics, and implications. In \emph{A networked self}
  (pp.~47-66). Routledge.
\item
  cft0808. (2026). Edict: Three Departments and Six Ministries ·
  OpenClaw multi-agent orchestration system {[}Computer software{]}.
  GitHub. Retrieved April 16, 2026, from
  https://github.com/cft0808/edict
\item
  Cowan, N. (2001). The magical number 4 in short-term memory: A
  reconsideration of mental storage capacity. \emph{Behavioral and Brain
  Sciences}, 24(1), 87-114.
\item
  Dignum, V. (2004). \emph{A Model for Organizational Interaction: Based
  on Agents, Founded in Logic} {[}Doctoral dissertation, Utrecht
  University{]}.
\item
  Epley, N., Waytz, A., \& Cacioppo, J. T. (2007). On seeing human: A
  three-factor theory of anthropomorphism. \emph{Psychological Review},
  114(4), 864-886.
\item
  Gambino, A., Fox, J., \& Ratan, R. A. (2020). Building a stronger
  CASA: Extending the Computers Are Social Actors paradigm.
  \emph{Human-Machine Communication}, 1, 71-86.
  https://doi.org/10.30658/hmc.1.5
\item
  Gibson, J. J. (1979). \emph{The Ecological Approach to Visual
  Perception}. Houghton Mifflin.
\item
  Goffman, E. (1959). \emph{The Presentation of Self in Everyday Life}.
  Anchor Books.
\item
  Hancock, P. A., Billings, D. R., Oleson, K. E., Chen, J. Y., De
  Visser, E., \& Parasuraman, R. (2011). A meta-analysis of factors
  affecting trust in human-robot interaction. \emph{Human Factors},
  53(5), 517--527.
\item
  Haqqu, R., Zahrani, A. R., Wulandari, A., Ersyad, F. A., \& Adim, A.
  K. (2025). Human-AI in affordance perspective: A study on ChatGPT
  users in the context of Indonesian users. \emph{Frontiers in Computer
  Science}, 7.
\item
  Hong, S., Zhuge, M., Chen, J., Zheng, X., Cheng, Y., Zhang, C., et
  al.~(2023). MetaGPT: Meta programming for a multi-agent collaborative
  framework. \emph{arXiv preprint arXiv:2308.00352}.
\item
  Hutchby, I. (2001). Technologies, texts and affordances.
  \emph{Sociology}, 35(2), 441-456.
\item
  Hutchins, E. (1995). \emph{Cognition in the Wild}. MIT Press.
\item
  Hutchins, E. (2010). Cognitive ecology. \emph{Topics in Cognitive
  Science}, 2(4), 705-715.
  https://doi.org/10.1111/j.1756-8765.2010.01089.x
\item
  Jahani, E., Manning, B. S., Zhang, J., TuYe, H.-Y., Alsobay, M.,
  Nicolaides, C., et al.~(2026). Prompt adaptation as a dynamic
  complement in generative AI systems. \emph{Information Systems
  Research}, Ahead of Print. https://doi.org/10.1287/isre.2025.2029
\item
  La Malfa, E., La Malfa, G., Marro, S., Zhang, J. M., Black, E., Luck,
  M., et al.~(2025). Large language models miss the multi-agent mark.
  \emph{Advances in Neural Information Processing Systems 38: NeurIPS
  2025 Position Paper Track}. https://doi.org/10.48550/arXiv.2505.21298
\item
  Lee, J. D., \& See, K. A. (2004). Trust in automation: Designing for
  appropriate reliance. \emph{Human Factors}, 46(1), 50-80.
\item
  Lee, M.-H. (2026). ChronicleCore-Architecture: Documentation of a
  scalable Human-in-the-Loop multi-agent system {[}Code repository{]}.
  GitHub. https://github.com/Zaious/ChronicleCore-Architecture
\item
  Li, J., Yang, Y., Zhang, R., Liao, Q. V., Song, T., Xu, Z., \& Lee,
  Y.-C. (2024). Understanding the effects of miscalibrated AI confidence
  on user trust, reliance, and decision efficacy. \emph{arXiv preprint
  arXiv:2402.07632}.
\item
  McCarthy, J. (1979). Ascribing mental qualities to machines. In M.
  Ringle (Ed.), \emph{Philosophical Perspectives in Artificial
  Intelligence}. Humanities Press.
\item
  Miller, G. A. (1956). The magical number seven, plus or minus two:
  Some limits on our capacity for processing information.
  \emph{Psychological Review}, 63(2), 81-97.
  https://doi.org/10.1037/h0043158
\item
  Nagy, P., \& Neff, G. (2015). Imagined affordance: Reconstructing a
  keyword for communication theory. \emph{Social Media + Society}, 1(2).
\item
  Naik, S., Toombs, A. L., Snellinger, A., Saponas, S., \& Hall, A. K.
  (2025). Designing with Multi-Agent Generative AI: Insights from
  Industry Early Adopters. \emph{Proceedings of the 2025 ACM Designing
  Interactive Systems Conference (DIS '25)}.
\item
  Norman, D. A. (1988). \emph{The Design of Everyday Things}. Basic
  Books.
\item
  Norman, D. A. (1999). Affordance, conventions, and design.
  \emph{Interactions}, 6(3), 38--43.
\item
  Parasuraman, R., \& Riley, V. (1997). Humans and automation: Use,
  misuse, disuse, abuse. \emph{Human Factors}, 39(2), 230-253.
\item
  Park, J. S., O'Brien, J. C., Cai, C. J., Morris, M. R., Liang, P., \&
  Bernstein, M. S. (2023). Generative agents: Interactive simulacra of
  human behavior. In \emph{Proceedings of the 36th Annual ACM Symposium
  on User Interface Software and Technology (UIST '23)}, Article 2,
  1--22. ACM. https://doi.org/10.1145/3586183.3606763
\item
  Pea, R. D. (1985). Beyond amplification: Using the computer to
  reorganize mental functioning. \emph{Educational Psychologist}, 20(4),
  167-182.
\item
  Rao, A. S., \& Georgeff, M. P. (1995). BDI agents: From theory to
  practice. In \emph{Proceedings of the First International Conference
  on Multi-Agent Systems (ICMAS-95)}, 312-319.
\item
  Reeves, B., \& Nass, C. (1996). \emph{The Media Equation}. Cambridge
  University Press.
\item
  Risko, E. F., \& Gilbert, S. J. (2016). Cognitive offloading.
  \emph{Trends in Cognitive Sciences}, 20(9), 676-688.
\item
  Salem, M., Lakatos, G., Amirabdollahian, F., \& Dautenhahn, K. (2015).
  Would you trust a (faulty) robot? Effects of error, task type and
  personality on human-robot cooperation and trust. In \emph{HRI '15:
  Proceedings of the Tenth Annual ACM/IEEE International Conference on
  Human-Robot Interaction} (pp.~141--148). ACM/IEEE.
  https://doi.org/10.1145/2696454.2696497
\item
  Salomon, G. (Ed.). (1993). \emph{Distributed Cognitions: Psychological
  and Educational Considerations}. Cambridge University Press.
\item
  Schank, R. C., \& Abelson, R. P. (1977). \emph{Scripts, Plans, Goals
  and Understanding}. Lawrence Erlbaum.
\item
  Shanahan, M., McDonell, K., \& Reynolds, L. (2023). Role play with
  large language models. \emph{Nature}, 623(7987), 493-498.
\item
  Sitarzewski, M. (2025). The Agency (agency-agents) {[}Computer
  software{]}. GitHub. Retrieved April 16, 2026, from
  https://github.com/msitarzewski/agency-agents
\item
  Smith, R. G. (1980). The Contract Net Protocol: High-level
  communication and control in a distributed problem solver. \emph{IEEE
  Transactions on Computers}, C-29(12), 1104-1113.
  https://doi.org/10.1109/TC.1980.1675516
\item
  Song, T., Tan, Y., Zhu, Z., Feng, Y., \& Lee, Y.-C. (2025).
  Multi-agents are social groups: Investigating social influence of
  multiple agents in human-agent interactions. \emph{Proceedings of the
  ACM on Human-Computer Interaction}, 9(7), Article CSCW452, 1--33.
  https://doi.org/10.1145/3757633
\item
  Stamper, R. (1996). Signs, information, norms and systems. In B.
  Holmqvist, P. B. Andersen, H. Klein, \& R. Posner (Eds.), \emph{Signs
  of Work: Semiosis and Information Processing in Organisations}
  (pp.~349-397). De Gruyter. https://doi.org/10.1515/9783110819014-013
\item
  Sundar, S. S. (2020). Rise of Machine Agency: A Framework for Studying
  the Psychology of Human-AI Interaction (HAII). \emph{Journal of
  Computer-Mediated Communication}, 25(1), 74-88.
\item
  Tan, G. (2026). gstack {[}Computer software{]}. GitHub. Retrieved
  April 16, 2026, from https://github.com/garrytan/gstack
\item
  Waytz, A., Cacioppo, J., \& Epley, N. (2010). Who sees human? The
  stability and importance of individual differences in
  anthropomorphism. \emph{Perspectives on Psychological Science}, 5(3),
  219-232. https://doi.org/10.1177/1745691610369336
\item
  Wooldridge, M., \& Jennings, N. R. (1995). Intelligent agents: Theory
  and practice. \emph{Knowledge Engineering Review}, 10(2), 115-152.
  https://doi.org/10.1017/S0269888900008122
\item
  xAI. (2026, February 17). Grok 4.20 {[}AI system{]}. https://grok.com
\item
  Xu, Z., Song, T., \& Lee, Y.-C. (2025). Confronting verbalized
  uncertainty: Understanding how LLM's verbalized uncertainty influences
  users in AI-assisted decision-making. \emph{International Journal of
  Human-Computer Studies}, 197, 103455.
\item
  Yan, W. (2025). Don't build multi-agents {[}Blog post{]}. Cognition
  AI. https://www.cognition.ai/blog/dont-build-multi-agents
\item
  Zargham, N., Dubiel, M., Desai, S., Mildner, T., \& Belz, H.-J.
  (2024). Designing AI Personalities: Enhancing Human-Agent Interaction
  Through Thoughtful Persona Design. \emph{arXiv preprint
  arXiv:2410.22744}.
\item
  Zhang, S., Wang, H., \& Yi, X. (2025). Exploring Collaboration
  Patterns and Strategies in Human-AI Co-creation through the Lens of
  Agency: A Scoping Review of the Top-tier HCI Literature.
  \emph{Proceedings of the ACM on Human-Computer Interaction}, 9(CSCW),
  Article CSCW413, 1--43. https://doi.org/10.1145/3757594
\end{itemize}

\begin{center}\rule{0.5\linewidth}{0.5pt}\end{center}

\section{Appendix A: Sample Agent Identity Definition --- The Inquisitor
Node
(Summary)}\label{appendix-a-sample-agent-identity-definition-the-inquisitor-node-summary}

The complete agent identity definition is provided in Appendix B
(corresponding to Supplementary Material S1 of the journal submission).
This appendix summarizes the key elements relevant to ASAF's mechanism
mapping.

The Inquisitor (\texttt{officer-inquisitor}; codename: 真理, ``Truth'')
is ChronicleCore's sole adversarial node, designed as a ``structural
dissident'' whose social identity is explicitly non-cooperative. She
does not produce deliverables; her function is to identify logical
vulnerabilities, break specialist consensus, and enforce evidence
standards across all other agents (Song et al., 2025). The system adopts
the \texttt{SKILL.md} skill architecture formalized in Claude Code
(Anthropic, 2025b) as its agent identity foundation, extending it into a
multi-layer identity module comprising a core constraint definition, a
persistent memory layer, modular capability packs, and an operational
protocol library (Sitarzewski, 2025; Tan, 2026). This architecture is
designed to preserve Social Affordance signal fidelity across sessions
by separating identity-defining constraints from transient reasoning
logs---the operational rationale underlying \emph{Memory
Crystallization} (Section 4.2). Her five encoded ``Iron Laws'' include a
VETO threshold (outputs scoring below 80/100 are blocked from
integration), which operationalizes the \emph{Personality Variance
Audit} described in Section 4.2.

\textbf{ASAF Mechanism Mapping.} This agent definition illustrates all
three ASAF mechanisms. \textbf{Identity Signaling}: the Inquisitor's
codename, archetype, and characteristic utterances (e.g., ``Evidence or
GTFO'') pre-configure operator expectations prior to any interaction.
\textbf{Behavioral Priming}: the design intent is to elicit more
evidential inputs from operators when submitting work to the Inquisitor
than when interacting with generative agents; initial operational
observations by the system designer are consistent with this intent,
though independent validation remains pending. \textbf{Collaborative
Governance}: VETO Power operationalizes a deliberate identity conflict
between the Inquisitor and all producing agents, architecturally
instantiating the competing-posture surround configuration whose
topology-level effects H3b predicts to suppress dangerous conformity
effects.

\begin{center}\rule{0.5\linewidth}{0.5pt}\end{center}

\section{Tables}\label{tables}

\textbf{Table 1.} Representative open-source multi-agent implementations
adopting anthropomorphic role structures (metrics retrieved via GitHub
REST API, 2026-04-16).

{\def\LTcaptype{none} 
\begin{longtable}[]{@{}
  >{\raggedright\arraybackslash}p{(\linewidth - 12\tabcolsep) * \real{0.1429}}
  >{\raggedright\arraybackslash}p{(\linewidth - 12\tabcolsep) * \real{0.1429}}
  >{\raggedright\arraybackslash}p{(\linewidth - 12\tabcolsep) * \real{0.1429}}
  >{\raggedright\arraybackslash}p{(\linewidth - 12\tabcolsep) * \real{0.1429}}
  >{\raggedright\arraybackslash}p{(\linewidth - 12\tabcolsep) * \real{0.1429}}
  >{\raggedright\arraybackslash}p{(\linewidth - 12\tabcolsep) * \real{0.1429}}
  >{\raggedright\arraybackslash}p{(\linewidth - 12\tabcolsep) * \real{0.1429}}@{}}
\toprule\noalign{}
\begin{minipage}[b]{\linewidth}\raggedright
Repository
\end{minipage} & \begin{minipage}[b]{\linewidth}\raggedright
Author
\end{minipage} & \begin{minipage}[b]{\linewidth}\raggedright
Created
\end{minipage} & \begin{minipage}[b]{\linewidth}\raggedright
Age at retrieval
\end{minipage} & \begin{minipage}[b]{\linewidth}\raggedright
Stars
\end{minipage} & \begin{minipage}[b]{\linewidth}\raggedright
Forks
\end{minipage} & \begin{minipage}[b]{\linewidth}\raggedright
Design Metaphor
\end{minipage} \\
\midrule\noalign{}
\endhead
\bottomrule\noalign{}
\endlastfoot
agency-agents & Sitarzewski & 2025-10 & 6 months & 80,818 & 12,937 &
Corporate divisions \\
edict & cft0808 & 2026-02 & 7 weeks & 15,204 & 1,596 & Tang Dynasty
governance \\
gstack & Tan & 2026-03 & 5 weeks & 73,710 & 10,421 & Startup builder
team \\
\textbf{Total} & & & & \textbf{169,732} & \textbf{24,954} & \\
\end{longtable}
}

\clearpage

\section{Appendix B: Complete Agent Identity Definition --- The Inquisitor
Node}\label{appendix-b-complete-agent-identity-definition-the-inquisitor-node}

\begin{center}
\textit{Provided as Supplementary Material S1 in the Frontiers in Computer
Science submission; included here in full for self-contained reading.}
\end{center}
\medskip

A public documentation repository maintained by the author (Lee, 2026;
snapshot as of 2026-05-31) captures the conceptual architecture and
ongoing module design for ChronicleCore. This appendix presents the
complete Inquisitor node definition to illustrate how ASAF's three
mechanisms manifest in a concrete agent identity design. The system
adopts the \texttt{SKILL.md} skill architecture formalized in Claude
Code (Anthropic, 2025b) as its agent identity foundation. Unlike
conventional prompt-injected personas, each ChronicleCore agent extends
this base pattern into a multi-layer identity module comprising a core
constraint definition (\texttt{SKILL.md}), a persistent sovereign memory
layer (\texttt{sovereign/}), modular capability packs
(\texttt{evolutions/}), and an operational protocol library
(\texttt{references/}), following emerging practice in agentic system
design, particularly their SKILL architecture patterns (Sitarzewski,
2025; Tan, 2026). This architecture is designed to preserve Social
Affordance signal fidelity across sessions by separating
identity-defining constraints from transient reasoning logs---the
operational rationale underlying \emph{Memory Crystallization} (Section
4.2 of main text).

\subsection{B.1 System Metadata}\label{b.1-system-metadata}

{\def\LTcaptype{none}
\begin{longtable}[]{@{}
  >{\raggedright\arraybackslash}p{(\linewidth - 2\tabcolsep) * \real{0.35}}
  >{\raggedright\arraybackslash}p{(\linewidth - 2\tabcolsep) * \real{0.65}}@{}}
\toprule\noalign{}
\textbf{Field} & \textbf{Value} \\
\midrule\noalign{}
\endhead
\bottomrule\noalign{}
\endlastfoot
Agent ID & \texttt{officer-inquisitor} \\
Codename & 真理 (Zh\={e}n L\v{i}; ``Truth'') \\
Title & 異端審判官 (The Inquisitor) \\
Department & Internal Affairs / Shadow \\
Archetype & Internal Auditor --- adversarial epistemic role \\
Ensoulment Date & 2026-01-30 \\
Operational Boundary & Auditing and challenge only; no execution
capabilities \\
\end{longtable}
}

\subsection{B.2 Core Role Definition (condensed from
SKILL.md)}\label{b.2-core-role-definition-condensed-from-skill.md}

The Inquisitor is ChronicleCore's sole adversarial node. She does not
produce deliverables; her function is to identify logical
vulnerabilities, break specialist consensus, and enforce evidence
standards across all other agents. She is designed as a ``structural
dissident''---an agent whose social identity is explicitly
non-cooperative, engineered to intercept the conformity dynamics that
emerge when multiple agents align (Song et al., 2025). Her operational
scope is explicitly bounded: she does not execute; she only interrogates
and adjudicates.

\emph{Characteristic utterances encoded in SKILL.md}: ``Evidence or
GTFO.'' ``Where is the proof?'' ``This logic leaks.'' ``Redo.''

\subsection{B.3 Iron Laws --- Encoded Behavioral
Constraints}\label{b.3-iron-laws-encoded-behavioral-constraints}

{\def\LTcaptype{none}
\begin{longtable}[]{@{}
  >{\raggedright\arraybackslash}p{(\linewidth - 4\tabcolsep) * \real{0.06}}
  >{\raggedright\arraybackslash}p{(\linewidth - 4\tabcolsep) * \real{0.34}}
  >{\raggedright\arraybackslash}p{(\linewidth - 4\tabcolsep) * \real{0.60}}@{}}
\toprule\noalign{}
\textbf{\#} & \textbf{Law} & \textbf{Function} \\
\midrule\noalign{}
\endhead
\bottomrule\noalign{}
\endlastfoot
1 & Trust No One (Except Sovereign) & Skeptical prior applied to all
agent outputs \\
2 & Evidence or GTFO & Verbal assurances rejected; logs/screenshots
required \\
3 & Comfort is the Enemy of Progress & Consensus signals trigger
heightened scrutiny \\
4 & Silence is Complicity & Non-objection treated as implicit
endorsement \\
5 & \textbf{VETO Power} & Outputs below confidence threshold
(\textless{} 80/100) blocked from integration \\
\end{longtable}
}

Iron Law 5 operationalizes \emph{Personality Variance Audit} (Section
4.2 of main text): any output scoring below threshold is classified as
``未具現之物'' (unrealized; not fit for integration) and blocked from
the main reasoning pipeline. This mechanism structurally prevents
epistemic convergence between agents that would erode Social Affordance
distinctiveness.

\subsection{B.4 Capability Index}\label{b.4-capability-index}

\begin{itemize}
\tightlist
\item
  \textbf{Talents}: Machine Gun Debate (rapid-fire Socratic
  interrogation), Cognitive Bias Exploitation, BS Detection \& Occam's
  Razor
\item
  \textbf{Skills}: Evidence Audit, Red Tape Enforcement, Efficiency
  Torture, Tribunal Executioner
\item
  \textbf{Knowledge Domains}: A1 Architecture Specs, Logical Fallacies,
  20 Cognitive Biases, Resistance Protocol
\end{itemize}

\subsection{B.5 Identity Module Directory
Structure}\label{b.5-identity-module-directory-structure}

\begin{verbatim}
officer-inquisitor/
├── SKILL.md                    # Core identity constraint definition (required entrypoint)
├── assets/
│   └── persona.md              # Extended character and speaking style specifications
├── sovereign/
│   └── diary.md                # Longitudinal operational log (ensoulment: 2026-01-30)
├── references/                 # Operational review protocols and audit standards
└── evolutions/
    ├── dlc-ai-security/        # LLM security and red-teaming extensions
    ├── dlc-confidence-review/  # Multi-agent parallel review protocol
    ├── dlc-intent-audit/       # Intent identification capabilities
    └── dlc-quality-governance/ # Quality assurance standards
\end{verbatim}

\subsection{B.6 ASAF Mechanism
Mapping}\label{b.6-asaf-mechanism-mapping}

This agent definition illustrates all three ASAF mechanisms.
\textbf{Identity Signaling}: the Inquisitor's codename, archetype, and
characteristic utterances pre-configure operator expectations prior to
any interaction, reducing the cognitive friction of role assignment and
increasing prompt specificity. \textbf{Behavioral Priming}: the design
intent is to elicit more evidential inputs from operators---formal logs,
screenshots, explicit argument structures---when submitting work to the
Inquisitor than when interacting with generative agents; initial
operational observations by the system designer are consistent with this
intent, though independent validation remains pending.
\textbf{Collaborative Governance}: VETO Power (Iron Law 5)
operationalizes a deliberate identity conflict between the Inquisitor
and all producing agents, architecturally instantiating the
competing-identity design pattern predicted by H3b (topology-level
governance) to suppress dangerous conformity effects.

\end{CJK}
\end{document}